\documentclass[10pt,journal,compsoc]{IEEEtran}
\usepackage[colorlinks,urlcolor=blue,linkcolor=blue,citecolor=blue]{hyperref}

\usepackage{cite}
\usepackage{amsmath,amssymb,amsfonts}
\usepackage{graphicx}
\usepackage{microtype}

\usepackage[export]{adjustbox}
\usepackage{doi}
\usepackage{siunitx}
\usepackage{booktabs,tabularx,ragged2e}
\newcolumntype{C}{>{\centering\arraybackslash}X}

\begin{document}

\title{Multimodal Effective Permittivity Model for Metasurfaces Embedded in Layered Media}

\author{Christopher~T.~Howard, William~D.~Hunt, Kenneth~W.~Allen}

\markboth{MULTIMODAL EFFECTIVE PERMITTIVITY MODEL FOR METASURFACES EMBEDDED IN LAYERED MEDIA}{HOWARD {\itshape ET AL}}

\author{Christopher~T.~Howard, 
        William~D.~Hunt, 
        Kenneth~W.~Allen%
\IEEEcompsocitemizethanks{
    \IEEEcompsocthanksitem The authors are with the School of Electrical and Computer Engineering, Georgia Institute of Technology, Atlanta, GA, 30332-0250 USA.\protect\\
    E-mail:~\href{mailto:christopher.howard@gtri.gatech.edu}{christopher.howard@gtri.gatech.edu}.
    \IEEEcompsocthanksitem C.~T.~Howard and K.~W.~Allen are also with the Advanced Concepts Laboratory, Georgia Tech Research Institute, Atlanta, GA 30332-0866}}%


\IEEEtitleabstractindextext{%
\begin{abstract}
The dielectric layers surrounding a metasurface have a large impact on its frequency and angular response. The notion of effective permittivity captures this dependence by suggesting that a layered dielectric environment will perturb metasurface resonances in the same manner as an infinite environment of that effective permittivity. A model for effective permittivity is presented in this work. In contrast to previous empirically derived models, this multi-term model is a robust approximation based on the modal expansion of fields at the metasurface. A simple dipole metasurface array is simulated to demonstrate the accuracy of the model, which is shown to be improved over other models. Finally, we show that it is possible for dielectric layers to produce an effective permeability different than unity, which must be considered in conjunction with higher-order equivalent circuit modeling to effectively predict metasurface impedance variation.
\end{abstract}}


%
\maketitle

\section{Introduction}

\IEEEPARstart{T}{he} presence of a supporting substrate is generally a necessity in the design of any metasurface, particularly for those having unconnected metallic elements. Aside from the mechanical support provided by the dielectric layers of a metasurface or frequency-selective surface (FSS), these layers have been shown to also provide electric benefits, such as improving the angular stability of a desired frequency response \cite{munk2000}, reducing the periodicity of a metasurface array, and improving bandwidth characteristics \cite{callaghan1991}. However, the analysis of the effects produced by changing the media in which a metasurface resides can be hampered by the computational burden of assessing each change. Because the metasurface's interaction with surrounding media is determined by an infinite number of evanescent harmonics with different decay rates \cite{monni2007}, there is no simple way to accurately predict how changing the supporting substrate's permittivity or thickness will affect the frequency response of the metasurface.

A robust treatment of the problem is the modal expansion method, where from an estimation of the currents or fields at the metasurface plane, generated by either an analytic representation or a computational electromagnetics (CEM) model, it is possible to obtain the amplitude of as many harmonics as necessary to fully characterize the surface. This approach is rooted in the variational techniques developed by Schwinger \cite{schwinger1968} for obstacles and apertures in waveguides, which was expanded to provide a multimodal network representation of frequency-selective surfaces and gratings \cite{palocz1970,contu1983}. This formulation was later used as the theoretical basis for numerical methods such as the spectral iteration method \cite{tsao1982}, the generalized scattering matrix \cite{hall1988} and Method of Moments codes incorporating periodicity \cite{cwik1987,yoo2011}. More recently, there has been interest \cite{garcia-vigueras2012,rodriguez-berral2015,mesa2016,hum2017,astorino2018,baladi2021,you2024} in quasi-analytic modal expansion techniques, where a surface current profile is obtained from a full-wave CEM simulation and expanded into a series of plane waves, both propagating and evanescent. The impedance contribution of each mode to the total composite metasurface impedance can then be determined through an analytic transmission line model determined by the surrounding dielectric layers. Once a sufficiently large set of harmonic amplitudes have been calculated, the complete frequency response of the metasurface in any dielectric environment can be rapidly computed.

\begin{figure}
    \centering
    \includegraphics[width=3.49in]{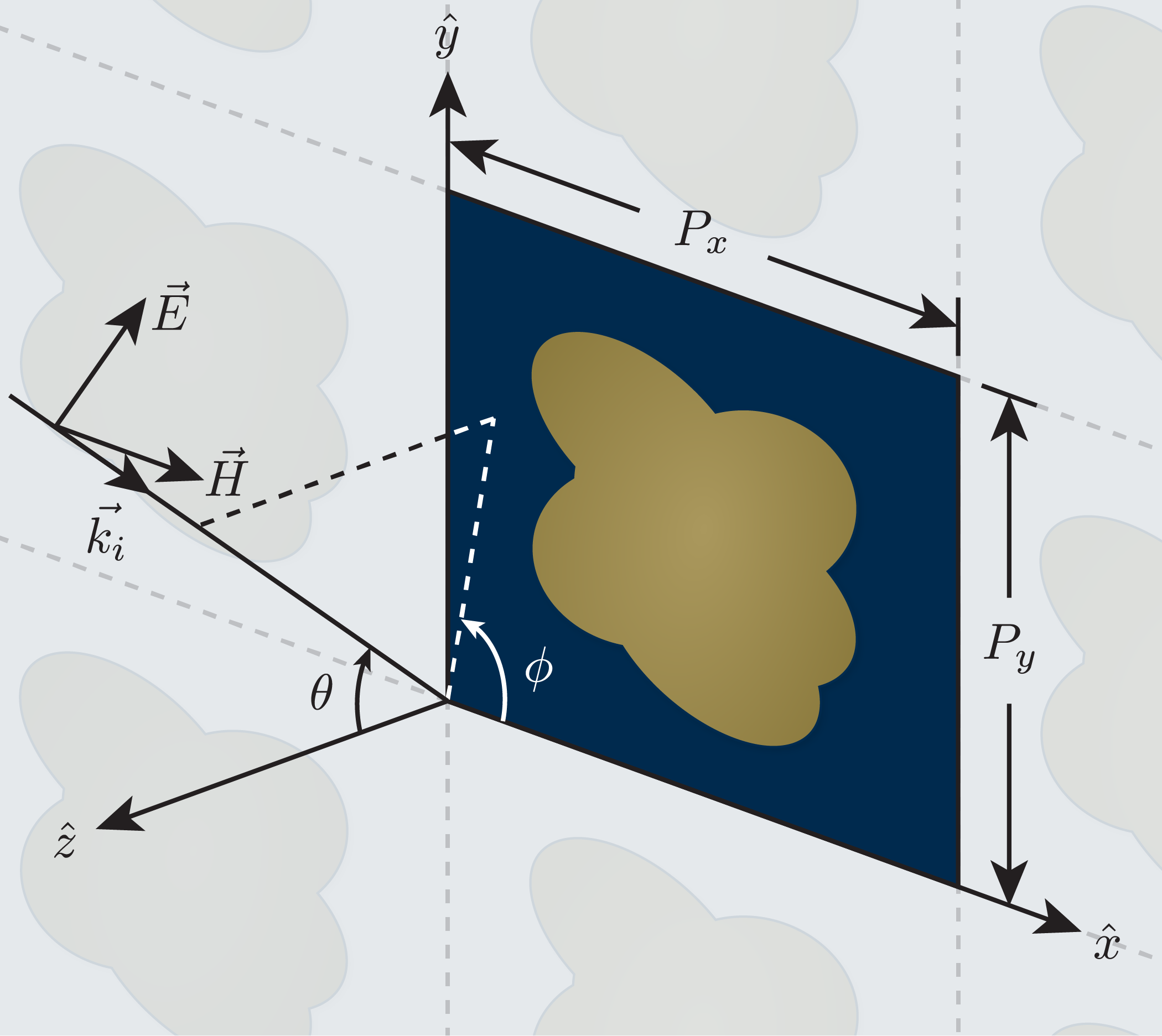}
    \caption{Dimensions for an arbitrary periodic surface and definitions for angle of incident waves.}
    \label{fig:periodic-floquet}
\end{figure}

The modal expansion technique, also known as the multimodal equivalent circuit method, is not without its shortcomings. Notably, there are underlying assumptions of the equivalent network which assume that the spatial current profile remains constant over the entire frequency region under examination, an assumption only true in a small set of canonical surfaces with only a single resonance. Efforts to extend the technique to cases where multiple independent spatial profiles and cross-polarization terms are underway \cite{conde-pumpido2022,hernandez-escobar2022,rodriguez-berral2022}, but even these cannot be applied in a general way to an arbitrary metasurface.

What is often desired is some estimation of how the dielectric layers will perturb a single dominant resonance in what may otherwise be a metasurface with a complex frequency response. It is well known that a dielectric layer immediately adjacent to a metasurface has the effect of shifting the resonant frequency downward; in terms of equivalent circuit terminology, this is because the capacitance of the surface is scaled by the \textit{effective permittivity} of the environment, which is a function of the complex permittivity of any nearby dielectric layers, the thickness of those layers, and the pattern of the metasurface itself. A simple exponential decay model describing this phenomenon has been presented by \cite{costa2021}, which will be used as a point of comparison throughout this work. In this paper, we present an improved model which recognizes that the effective permittivity with respect to thickness is dependent on multiple harmonic decays, rather than a single one, and computes effective permittivity based on an approximation of the rigorous modal expansion. This method is demonstrated to have substantially improved accuracy and predictive power over the single-term effective permittivity model.

\newcommand{\Zeq}{Z_{\text{eq}}}
\newcommand{\ereff}{\epsilon_{r,\text{eff}}}
\newcommand{\ureff}{\mu_{r,\text{eff}}}
\newcommand{\ereffh}{\epsilon_{rh,\text{eff}}}
\newcommand{\erinh}{\epsilon_{rh,\text{in}}}
\newcommand{\erone}{\epsilon_{r1}}
\newcommand{\ertwo}{\epsilon_{r2}}
\newcommand{\etah}[1]{\eta_{h}^{(#1)}}
\newcommand{\tankzh}[1]{\tan{k_{zh}^{(#1)}} d_{#1}}
\newcommand{\epsr}[1]{\epsilon_{r}^{(#1)}}
\newcommand{\tanad}[1]{\tan{(-j \alpha_{mn}^{(#1)}} d_{#1})}
\newcommand{\etaTMmn}{\eta_{\mathrm{TM}_{mn}}}
\newcommand{\etaTEmn}{\eta_{\mathrm{TE}_{mn}}}

\section{Multimodal Equivalent Circuit Modeling}

A metasurface can be represented by a two-port network. Equivalent circuit modeling (ECM) attempts to synthesize the impedance response of this two-port network into a circuit of lumped components \cite{costa2014}, often with an analytical relationship tying geometrical parameters to component values. The difficulty is in determining the arrangement of these elements as a function of the geometry. We now turn to a rigorous consideration of this problem.

\subsection{Formulation of the modal expansion}

As a consequence of Floquet theory \cite{floquet1883}, the fields on a periodic surface at $z=0$ can be decomposed into an infinite series of harmonics:
\begin{equation}
    E(x, y) = \sum_{h=0}^{\infty} A_{h} \bar{e}_h(x,y)
    \label{eq:plane-wave-basis}
\end{equation}  
where $\bar{e}_h$ are orthonormal basis vectors which represent spatial harmonics and must have the same periodicity as the field, and $A_h$ are coefficients describing the magnitude of each mode. The plane wave is a natural choice of basis:
\begin{equation}
    \bar{e}_h(x, y) = \frac{1}{\sqrt{P_xP_y}} e^{-j(k_{xm} x + k_{yn} y)} \hat{e}_h
\end{equation}
where $P_x$ and $P_y$ are the periodicity of the unit cell in orthogonal directions, as illustrated in \autoref{fig:periodic-floquet}. The harmonics must exist only in an infinite series of discrete modes with propagation direction dependent on the periodicity of the structure \cite{cwik1987}
\begin{align}
    k_{xm} &= \frac{2\pi m}{P_x} + k\sin{\theta_i}\cos{\phi_i}\\
    k_{yn} &= \frac{2\pi n}{P_y} + k\sin{\theta_i}\sin{\phi_i}
\end{align}
where  $k$ is the wavenumber in the incident media and $(\theta_i, \phi_i)$ is the angle of the incident wave (incidence angles other than normal will not be considered throughout this work, but are included here for completeness). For each mode it is possible to define a free-space cutoff frequency $\omega_{cmn}$ where propagation in the z-direction is zero ($k_z = 0$ for $\epsilon_r = \mu_r = 1$, where $\epsilon_r$ and $\mu_r$ are the relative permittivity and permeability, respectively) of
\begin{equation}
    \omega_{cmn} = \begin{cases}
        \phantom{2\pi}c\sqrt{k_{xm}^2 + k_{yn}^2} & \text{generally, or} \\
        2\pi c\sqrt{\left(\frac{m}{P_x}\right)^2 + \left(\frac{n}{P_y}\right)^2} 
            & \theta_i = 0
    \end{cases}
\end{equation}
where $c$ is the speed of light. Defined in terms of this free-space cutoff frequency, the wavenumber in the direction normal to the plane of the surface is 
\begin{equation}
    k_{zmn} = \begin{cases}
        -\dfrac{j}{c} \sqrt{\omega_{cmn}^2 - \epsilon_r \mu_r \omega^2}, & 
            \omega < \dfrac{\omega_{cmn}}{\sqrt{\epsilon_r\mu_r}} \\
        \phantom{-}\dfrac{1}{c} \sqrt{\epsilon_r\mu_r\omega^2 - \omega_{cmn}^2}, &
            \omega \geq \dfrac{\omega_{cmn}}{\sqrt{\epsilon_r\mu_r}}. \\
    \end{cases}
\end{equation}
Above the cutoff frequency, which is the free-space cutoff frequency scaled by the velocity of propagation in the medium, the wavenumber is real and therefore represents a propagating mode. Below cutoff, however, the wavenumber is imaginary, and $k_{zmn} = -j \alpha_{mn}$. These modes, which represent the vast majority of the modes in a surface, are evanescent; they do not propagate and their amplitude decays away from the surface as $\exp{(-\alpha_{mn} z)}$. Dividing the harmonics into transverse-electric (TE) and transverse-magnetic (TM) modes, the characteristic impedance of each evanescent wave is
\begin{align}
    \etaTMmn &= \dfrac{\alpha_{mn}}{j\omega\epsilon_0\epsilon_r} & \text{for TM modes} \label{eq:tm-impedance} \\
    \etaTEmn &= \dfrac{j\omega\mu_0\mu_r}{\alpha_{mn}} & \text{for TE modes} \label{eq:te-impedance}
\end{align}

The form of these equations is similar to the impedance of a capacitor $C$ and an inductor $L$, respectively
\begin{align}
    Z_C = \dfrac{1}{j\omega C} \\
    Z_L = j\omega L.
\end{align}
\begin{figure}
    \centering
    \includegraphics{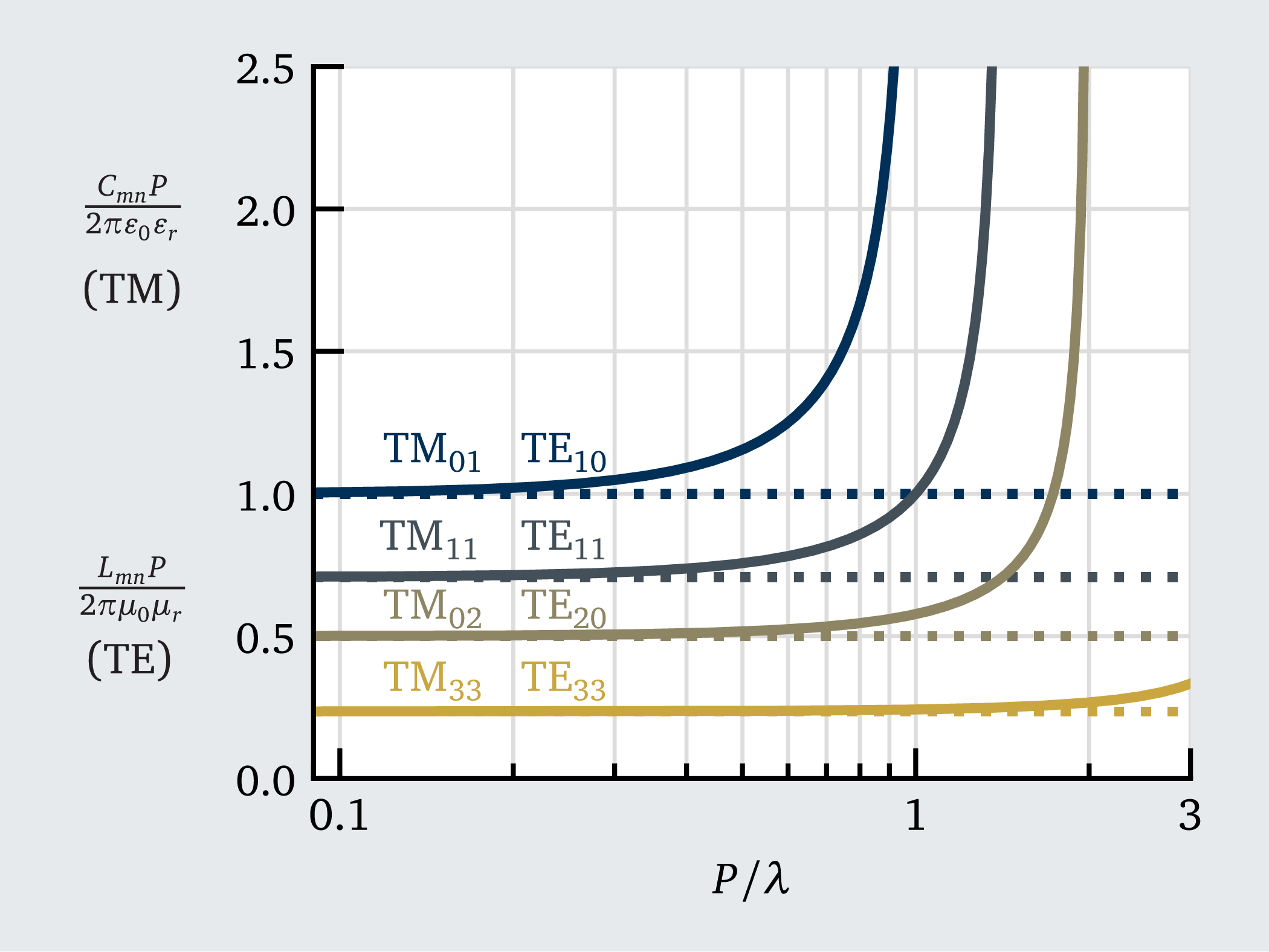}
    \caption{The impedance of evanescent harmonics on a periodic surface can be represented as equivalent capacitance (for TM modes) or inductance (for TE modes). At low frequencies, these values are static (the dashed line), but as the frequency approaches the cutoff frequency for the mode, they rapidly go to infinity (infinite impedance for TE modes, zero impedance for TM modes).}
    \label{fig:harmonic-lumped}
\end{figure}
Each evanescent TM harmonic can be represented as an equivalent capacitance 
\begin{equation}
    C_{mn}(\omega) = \frac{\epsilon_0\epsilon_r}{\alpha_{mn}(\omega)}
    \label{eq:modal-C}
\end{equation}
and each TE harmonic as an equivalent inductance 
\begin{equation}
    L_{mn}(\omega) = \frac{\mu_0\mu_r}{\alpha_{mn}(\omega)}.
    \label{eq:modal-L}
\end{equation}
This relationship is the basis for equivalent circuit modeling of periodic structures. Any ECM representation assumes static lumped component values -- the static values are themselves expressions of frequency-dependent impedance behaviors -- but the expressions for \eqref{eq:modal-C} and \eqref{eq:modal-L} vary considerably with frequency, as seen in \autoref{fig:harmonic-lumped}. However, at lower frequencies, $C_{mn}(\omega)$ and $L_{mn}(\omega)$ asymptotically approach a constant value such that they can be represented by a static capacitance and inductance, respectively:
\begin{align}
    C_{mn0} &= \frac{\epsilon_0\epsilon_r}{2\pi\sqrt{\left(\frac{m}{P_x}\right)^2 + \left(\frac{n}{P_y}\right)^2}}\label{eq:eff-inductance} \\
    L_{mn0} &= \frac{\mu_0\mu_r}{2\pi\sqrt{\left(\frac{m}{P_x}\right)^2 + \left(\frac{n}{P_y}\right)^2}}\label{eq:eff-capacitance}.
\end{align}
The general validity of the ECM approach is confined to those frequencies where the impedance response is sufficiently described by these static elements; that is, at wavelengths much longer than the periodicity. 

\subsection{Equivalent circuit for an obstacle-based metasurface}

\begin{figure}
    \includegraphics[width=2.5in,center]{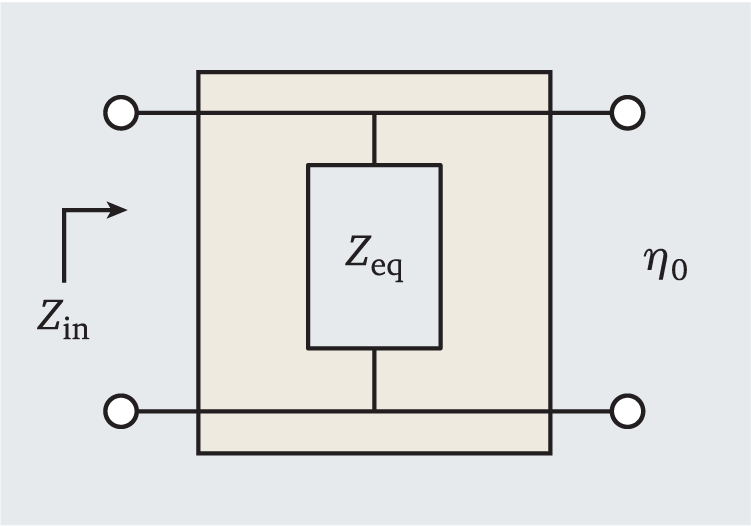}
    \caption{The equivalent two-port network of a freestanding metasurface has a shunt impedance representation.}
    \label{fig:network-shunt-impedance}
\end{figure}

The fact that individual TE and TM modes have equivalent representations as respectively inductors and capacitors says nothing about the nature of the circuit made up of those elements. In fact, the formulation of an equivalent circuit interaction among these harmonic static values is nontrivial and difficult to apply to an arbitrary metasurface which may have multiple independent current profiles \cite{schab2023}. However, Schwinger \cite{schwinger1968} in his development of variational techniques for analyzing waveguides determined that for a single polarization, a zero-thickness obstacle or aperture in a waveguide -- and consequently a metasurface -- must have a network consisting of a single shunt impedance $\Zeq$, as shown in \autoref{fig:network-shunt-impedance}; this arises somewhat intuitively from the continuity conditions between half-spaces on either side of the metasurface.
This network, which does not include the effects of any dielectric layers, has reflection coefficient
\begin{equation}
    S_{11} = \frac{\eta_0}{2\Zeq + \eta_0}
\end{equation}
and transmission
\begin{equation} 
    S_{21} = \frac{2\Zeq}{2\Zeq + \eta_0}
\end{equation}
where $\eta_0 = \sqrt{\mu_0/\epsilon_0}$ is the impedance of free space. Consequently, the surface impedance may be obtained from the reflection coefficient produced by a simulation or measurement with
\begin{equation}
    \Zeq = -\eta_0 \frac{1 + S_{11}}{2S_{11}}.
    \label{eq:impedance-from-S11}
\end{equation}
This impedance is purely imaginary when the surface is lossless (perfect conductivity) and induces no cross polarization components. For an obstacle-based metasurface with even symmetry and a constant spatial current profile over frequency, it has been shown in \cite{rodriguez-berral2015} that the equivalent impedance can be found by applying continuity conditions to \eqref{eq:plane-wave-basis} to find that
\begin{equation}
    Z_{eq} = \sum_{h'} A_h \eta_h
    \label{eq:simple-modal-sum}
\end{equation}
where $h'$ indicates the sum over the entire infinite series of harmonics except the incident harmonic and $A_h$ is the magnitude of each harmonic with modal impedance $\eta_h$ as defined by \eqref{eq:tm-impedance} and \eqref{eq:te-impedance}. The modal coefficients $A_h$ can be obtained from the spatial current profile $\mathbf{J}(x, y)$ by invoking the Fourier transform to obtain the spatial frequencies in both $x$ and $y$ for this current profile, denoted as $\widetilde{\mathbf{J}}_y(\mathbf{k}_{th})$ for $\mathbf{k}_{th} = k_{xm}\hat{x} + k_{yn}\hat{y}$:
\begin{equation}
    A_h = \left|\mathbf{\widetilde{J^*}}\left(\mathbf{k}_{th}\right) \cdot \hat{\mathbf{e}}_h \right|^2
\end{equation}
where
\begin{equation}
    \hat{\mathbf{k}}_{th} = \dfrac
        {k_{xm}\hat{\mathbf{x}} + k_{yn}\hat{\mathbf{y}}}
        {\sqrt{k_{xm}^2 + k_{yn}^2}}
\end{equation}
\begin{equation}
    \hat{\mathbf{e}}_h = \begin{cases}
        \hat{\mathbf{k}}_{th}, & \text{TM modes} \\
        \hat{\mathbf{k}}_{th} \times \hat{\mathbf{z}}, & \text{TE modes}.
    \end{cases}
\end{equation}

The spatial current profile $\mathbf{J}$ can be obtained from a single full-wave simulation or from some analytic representation of currents on the metasurface. The exact solution of \eqref{eq:simple-modal-sum} requires an infinite summation, but in a numerical solution this sum is truncated; for example, to the spatial resolution sampled by the full-wave solver.

\begin{figure}
    \centering
    \includegraphics[width=3.49in]{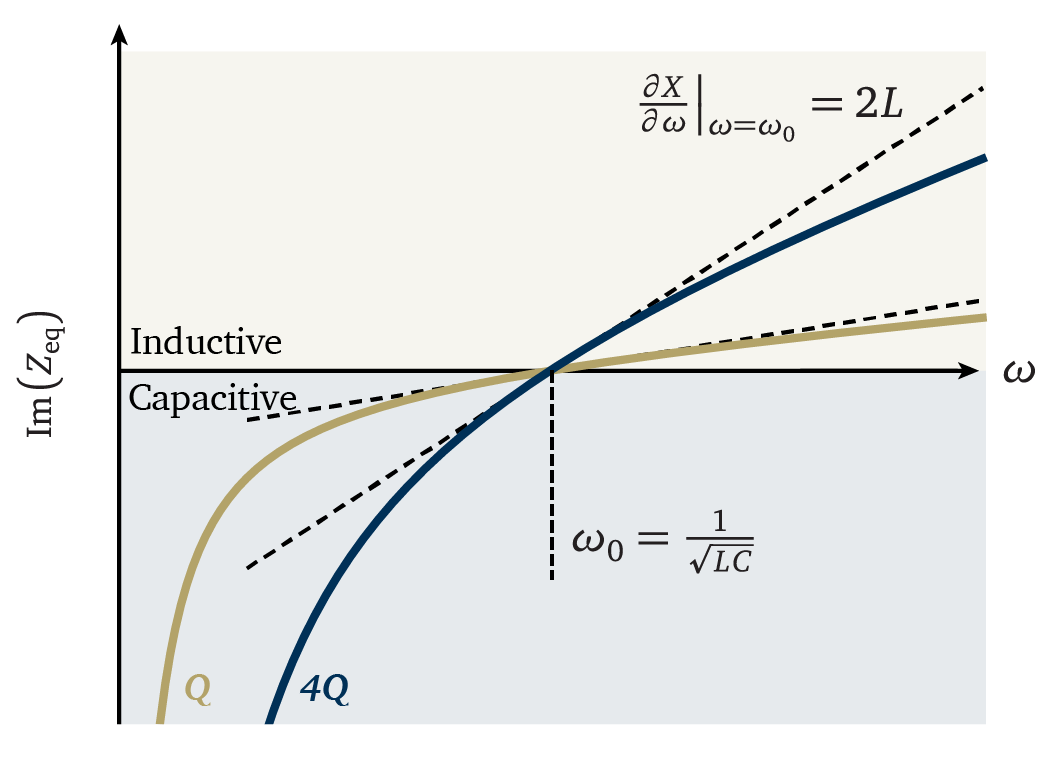}
    \caption{Reactance curves for series LC resonators. Resonance occurs when reactance is zero, and equivalent inductance can be computed from the slope at resonance.}
    \label{fig:notional-reactance}
\end{figure}

We have previously noted that TM modes behave as capacitors and TE modes as inductors, subject to a lack of dispersion in the modes. Considering TM and TE modes separately, the surface impedance may be represented as
\begin{equation}
    Z_{eq} = \frac{1}{j\omega C} + j\omega L
    \label{eq:simple-LC}
\end{equation}
with
\begin{align}
    \frac{1}{C} &= \sum_{h'}^{\text{TM}} \frac{A_h}{C_h} \\
    L &= \sum_{h'}^\text{TE} A_h L_h
\end{align}
where $C_h$ and $L_h$ are the modal equivalent values described by \eqref{eq:eff-capacitance} and \eqref{eq:eff-inductance}, respectively, and the summation over $h'$ is performed over every harmonic mode except the incident one. This equivalent circuit representation describes an LC circuit with an inductor and capacitor in series. The reactance curve for two different sets of component values is shown in \autoref{fig:notional-reactance}. 

The resonant frequency of this circuit -- the frequency at which the reactance is zero -- corresponds to a transmission null or reflection peak, when
\begin{equation}
    \omega_0 = \frac{1}{\sqrt{LC}}
\end{equation}
At resonance, the slope of the reactance is 
\begin{equation}
    \left.\frac{\partial}{\partial \omega} X_{eq} \right|_{\omega=\omega_0} = 2L
\end{equation}
which means that from a reactance curve obtained from either measurement or simulation, it is possible to calculate the equivalent inductance of the metasurface resonance as
\begin{equation}
    L = \frac{1}{2} \left. \frac{\partial X_{eq}}{\partial \omega} \right|_{\omega=\omega_0}.
    \label{eq:L-from-slope}
\end{equation}
The equivalent capacitance may then be computed as
\begin{equation}
    C=\frac{1}{\omega_0^2 L}.
    \label{eq:C-from-slope}
\end{equation}

\subsection{Effect of dielectric layers in the modal expansion}

\begin{figure}
    \centering
    \includegraphics[width=3.49in]{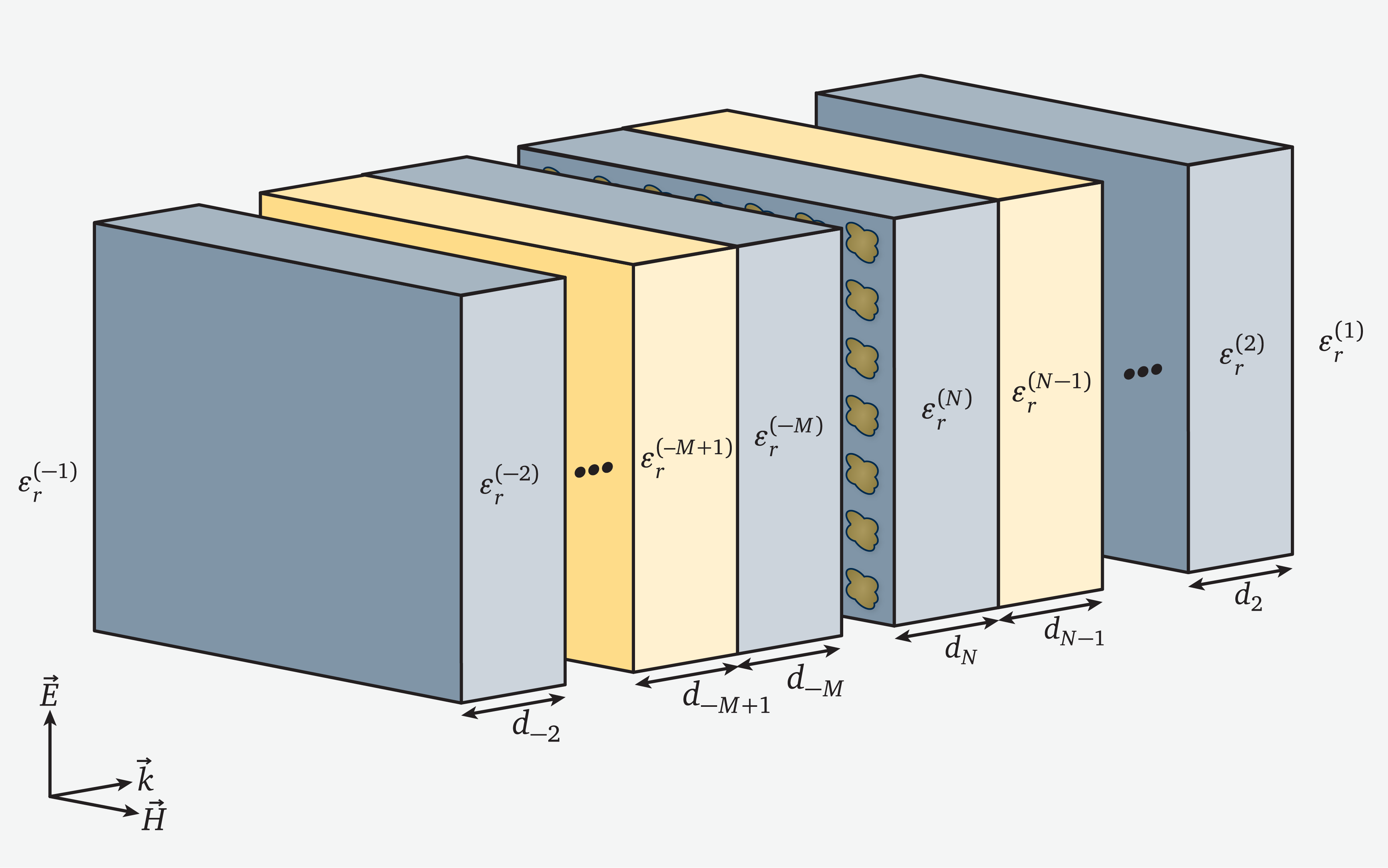}
    \caption{An arbitrary stackup with a metasurface sandwiched between $M$ dielectrics with permittivities $\varepsilon_{r}^{(i)}$ and thicknesses $d_i$ to the left, and $N$ dielectric layers to the right.}
    \label{fig:multilayer-stackup}
\end{figure}

\begin{figure*}
    \centering
    \includegraphics[width=6in]{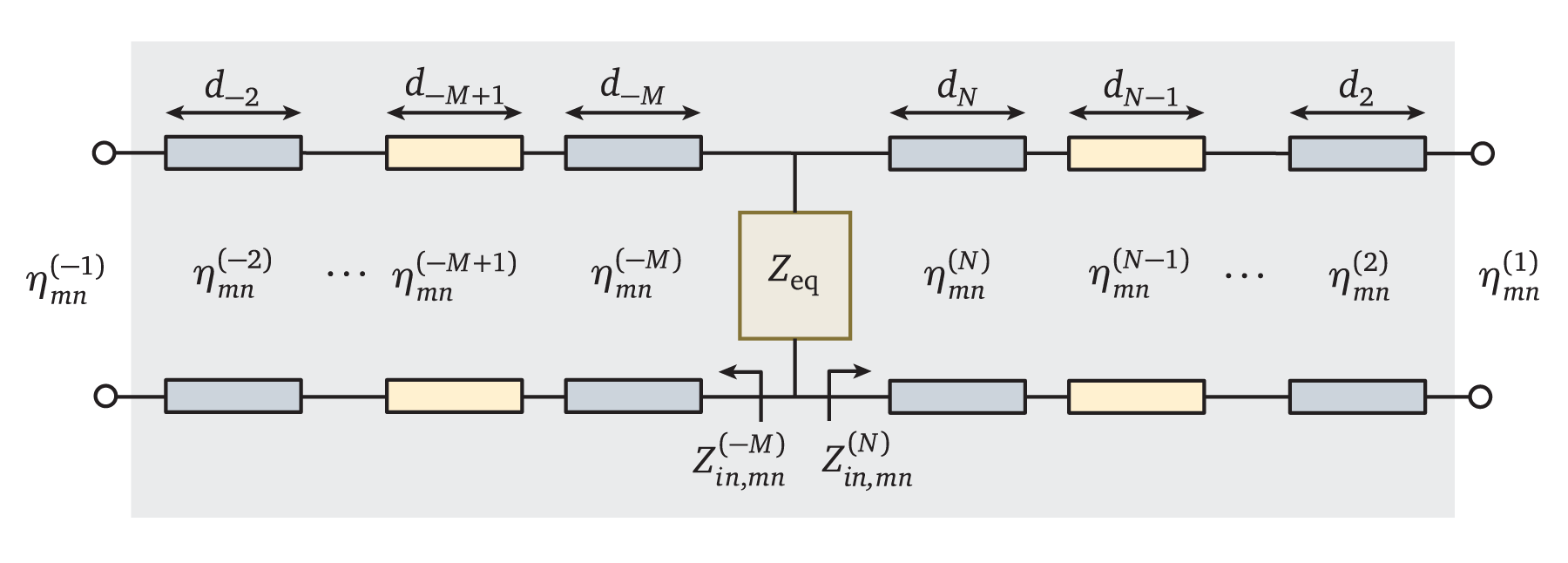}
    \caption{The layered stackup of \autoref{fig:multilayer-stackup} can be represented as a cascaded transmission line problem with metasurface as equivalent shunt impedance $\Zeq$.}
    \label{fig:multilayer-tlines}
\end{figure*}

Consider the arbitrary dielectric stackup illustrated in \autoref{fig:multilayer-stackup}, which consists of $N+M-2$ dielectric layers, and the $i$-th layer has permittivity $\epsilon_r^{(i)}$ and thickness $d_{i}$. This can be posed as the transmission line network shown in \autoref{fig:multilayer-tlines}; the shunt admittance of the surface is now surrounded on each side by transmission lines of length $d_{i}$ and characteristic impedance $\eta_{h}^{(i)}$ for each layer, terminated to the left by an impedance $\eta_{h}^{(-1)}$ and to the right by $\eta_{h}^{(1)}$. Each harmonic impedance in \eqref{eq:simple-modal-sum} is modified to the inverse of the sum of admittances looking to the left $Y_{in,h}^{(-M)}$ and to the right $Y_{in,h}^{(N)}$, so that the surface impedance is
\begin{equation}
    \Zeq = \sum_{h'} \frac{A_h}{Y_{in,h}^{(-M)} + Y_{in,h}^{(N)}}
    \label{eq:Zeq-sum}
\end{equation}
where $Y=1/Z$. The modal impedance looking into the $i$-th layer is given by the recursive transmission line equation
\begin{equation}
    Z_{\text{in,h}}^{(i)} = \etah{i} \frac
        {Z_{\text{in,h}}^{(i-1)} + j\etah{i} \tankzh{i}}
        {\etah{i} + j Z_{\text{in,h}}^{(i-1)} \tankzh{i}}
        \label{eq:transmission-line}
\end{equation}
When $k_{zh}$ in the layer is real, propagation is supported and the input impedance varies considerably and periodically with frequency or layer thickness. When $k_{zh}$ is imaginary, the mode is evanescent and decays with  the thickness of the material. In the evanescent case, $Z_{\text{in,h}}^{(i)}$ approaches $Z_{\text{in,h}}^{(i-1)}$ as the layer thickness $d_i$ approaches zero, and to the layer impedance $\etah{i}$ at large values of $d_i$.

\subsection{Validation of multimodal solution with CEM}

\begin{figure}
    \centering
    \includegraphics[width=2in]{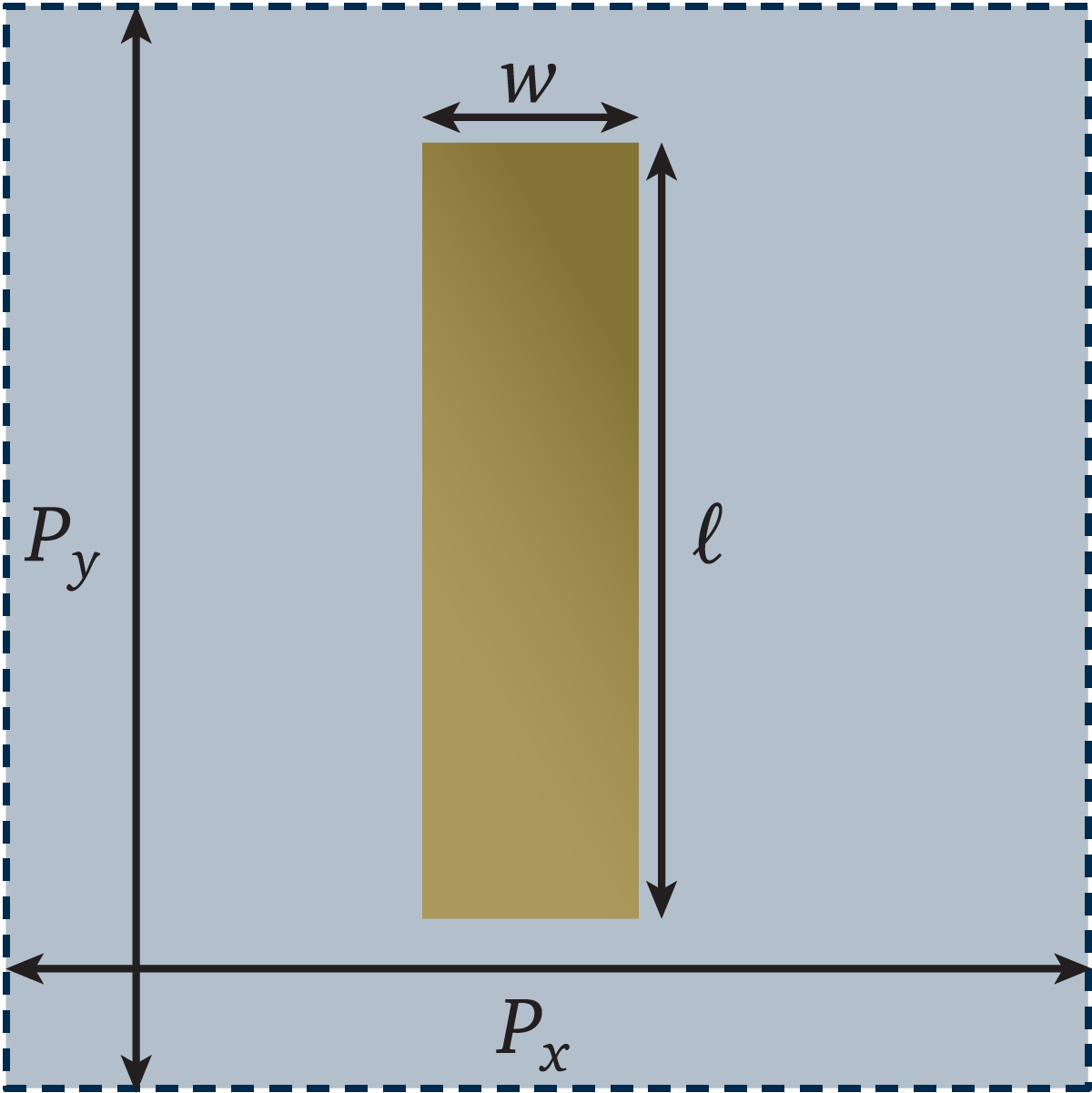}
    \caption{Dimensions of the periodic dipole metasurface simulated throughout this work, with length $\ell = \SI{9}{\mm}$, width $w=\SI{0.25}{\mm}$, and periodicity $P_x = P_y =\SI{10}{\mm}$ (other periodicities are also considered later).}
    \label{fig:dipole-dimensions}
\end{figure}

In order to evaluate the accuracy of the forthcoming effective permittivity model, it is necessary to compare any predictions from it against those obtained some other way, such as from a full-wave solver. A simple dipole array is simulated for this purpose, with a unit cell and dimensions shown in \autoref{fig:dipole-dimensions}.

A solution for the surface impedance is obtained via the analytic modal expansion approach described in \cite{rodriguez-berral2015} and \cite{baladi2021}, which is capable of analytically producing surface impedance assuming a particular spatial current profile. For the analytic model, an estimate of surface current is used for the modal expansion \cite{rodriguez-berral2013}:
\begin{equation}
    \mathbf{J} = \hat{y} \sqrt{\frac{1 - (2y / l)^2}{1 - (2x/w)^2}} \;\textrm{rect}\left(\frac{y}{l}\right) \;\textrm{rect}\left(\frac{x}{w}\right).
\end{equation}

\begin{figure*}
    \centering
    \includegraphics[width=6in]{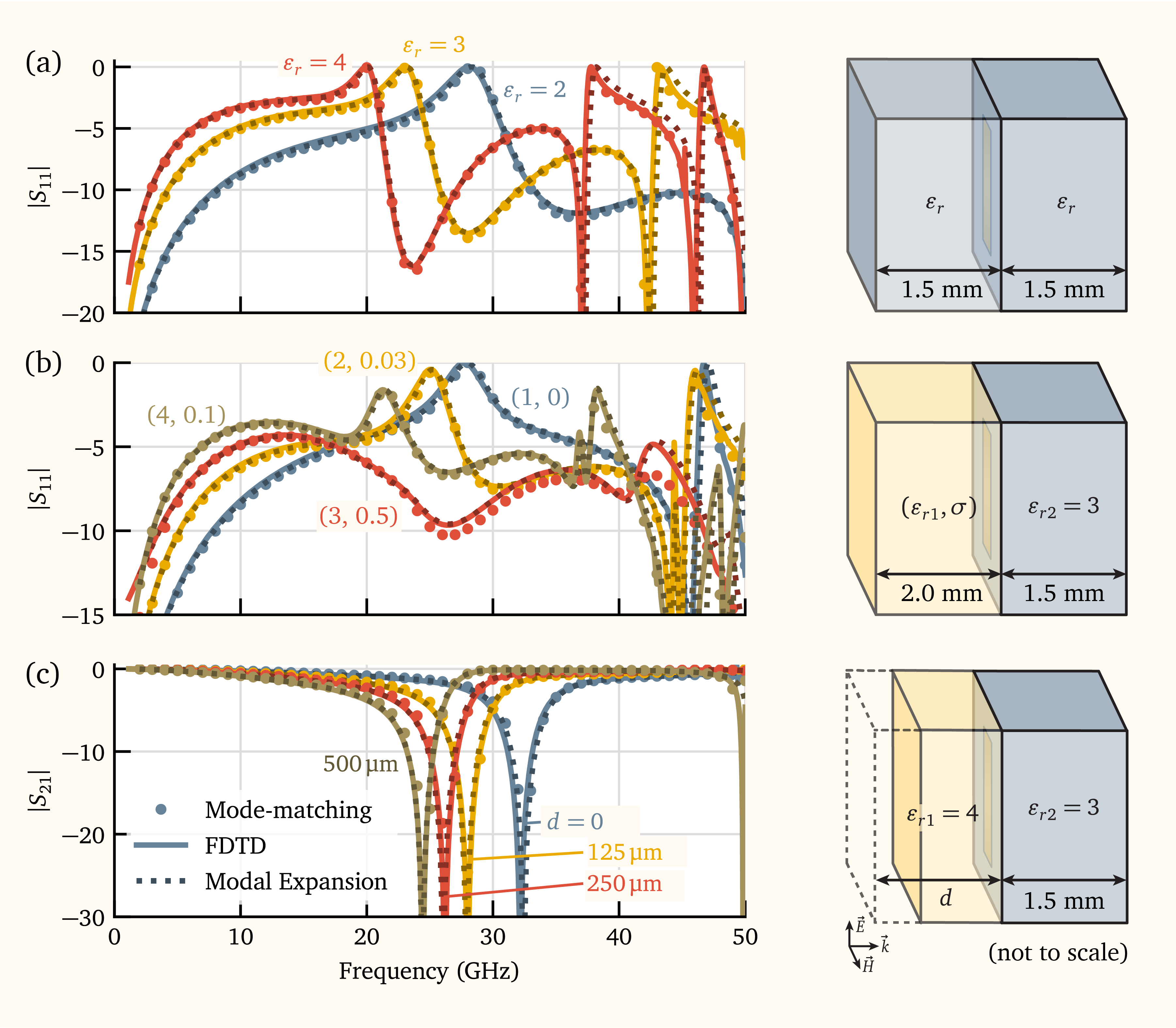}
    \caption{The accuracy of the modal expansion analysis (dashed lines) for calculating scattering from a dipole array is compared to both FDTD (solid lines) and a mode-matching code (circle markers). (a) Reflection for varying permittivity of identical substrates on either side of the dipole array. (b) Reflection for different substrates, one of which is lossy (conductivity $\sigma$ is used instead of loss tangent $\tan\delta$ for causality considerations in the FDTD solution). (c) Transmission with variation in the thickness of a thin substrate on one side of the metasurface.}
    \label{fig:modal-cem-comparison}
\end{figure*}

One advantage of this analytic model is that as part of its modal expansion it computes a large number of the harmonic amplitudes $A_h$, which are identical to those required to directly compute effective permittivity, allowing validation of the derived relationship and also detailed exploration of the nature of contributions from each harmonic towards the overall effective permittivity. One may object to the use of the modal expansion method to confirm the effectiveness of an effective permittivity model which presupposes the validity of the modal expansion method. To obviate this concern, comparison of results obtained from the modal expansion to both finite-difference time-domain (FDTD) results and those from a mode-matching code is shown in \autoref{fig:modal-cem-comparison}.

\section{Modal Effective Permittivity Model}

A metasurface layer -- an infinitesimally thin one -- suspended in free space with a periodic unit cell and resonant frequency $\omega_0$ will, when embedded in an infinitely thick dielectric environment of relative permittivity $\epsilon_r$, have a new resonant frequency 
\begin{equation}
    \omega_1 = \omega_0 \left/  \sqrt{\epsilon_r}\right.
\end{equation}
and indeed the entire frequency response will be scaled by such a factor as a direct consequence of Maxwell's equations; the propagation velocity has been slowed by a factor of $\sqrt{\epsilon_r}$. 
It is to be expected that if the metasurface were instead embedded between two dielectric layers of \textit{finite} thickness in free space, the resonant frequency would lie somewhere between $\omega_0$ and $\omega_1$; the limit of dielectric thickness approaching zero appears as free space, while increasing thickness eventually is indistinguishable from the dielectric half-space case. The concept of \textit{effective permittivity}, denoted by $\ereff$, describes the behavior of a particular metasurface between these two limits, suggesting that the impedance of a metasurface embedded in an arbitrary arrangement of dielectric layers will have an impedance identical to that of one embedded in an infinite medium of permittivity $\ereff$. We must however be careful to discriminate between changes in effective capacitance from changes in effective inductance; the character of the resonance in the frequency response depends on this distinction. We can therefore define both effective permittivity $\ereff$ and effective permeability $\ureff$ as
\begin{align}
    \ereff &= \frac{C_1}{C_0} \\
    \ureff &= \frac{L_1}{L_0}.
\end{align}
Because we are examining the impact of dielectric layers, effective permittivity is expected to dominate the impact on the metasurface. We may now turn to an examination of models for representing effective permittivity behavior.

\subsection{Single-term effective permittivity model}


One influential approximation of the behavior is that in \cite{costa2021} which represents the effective permittivity by a single-parameter exponential decay model:
\begin{equation}
    \ereff = \epsr{1} + \left(\epsr{2} - \epsr{1}\right) e^{-\alpha d_2 / P}.
    \label{eq:costa-model}
\end{equation}
This model has begun to see adoption in the literature \cite{kong2024,parameswaran2023,lestini2023,chang2022}, and therefore serves as a useful point of comparison for the model that follows. It shows that the \textit{shape factor} $\alpha$ is related to the fill factor of metal in the unit cell, which can be tied to the decay rate of underlying harmonic amplitudes. The model also highlights an important and perhaps non-intuitive point; nowhere in \eqref{eq:costa-model} is there a term for the frequency of operation of the metasurface. It is natural to assume that the effective permittivity convergence would be a function of the electric thickness of the dielectric layer; that is, of the wavelength.  
Any association with the frequency is strictly coincidental, and occurs because the design frequency for metasurfaces is itself often correlated with periodicity. Effective permittivity in reality is unrelated to the frequency of operation; in fact, the underlying assumption of an equivalent circuit model is a static capacitance entirely independent of frequency, which we should likewise expect from our effective permittivity model.

The merit of the single-term model is its ease of applicability, as it can capably approximate the thickness-dependent behavior over the range of common substrate thicknesses used for frequency-selective surfaces (on the order of millimeters). It is however inherently limited in the accuracy it can provide over several decades of thickness variation. The model, which is an experimentally-derived fit, represents the effective permittivity of only a single exponential decay term. This is equivalent to the effective permittivity of a single harmonic, and does not account for the effective permittivity as a process governed by a large number of harmonics with varying decay rates, the effect of which is to change the slope of convergence on a log scale.

\subsection{Effective permittivity from the modal expansion}

\begin{figure}
    \includegraphics[center,width=3.49in]{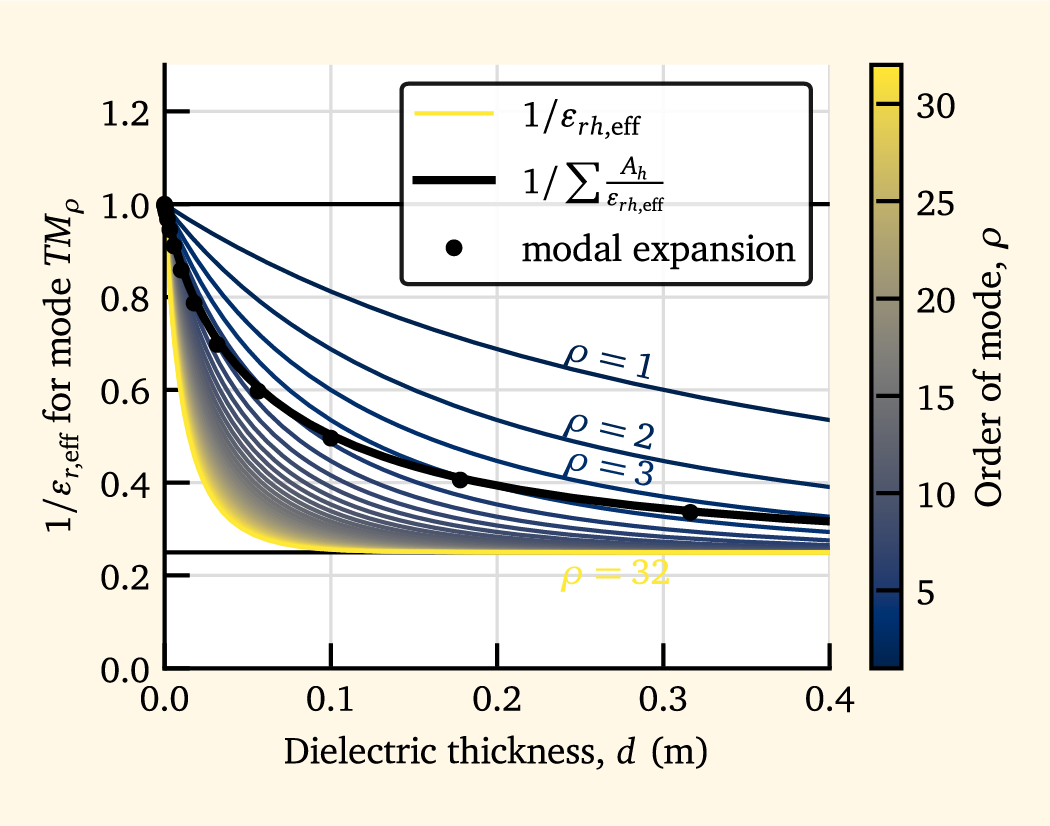}
    \caption{Effective permittivity looking into a dielectric layer of thickness for $d$ for $P=\SI{10}{\mm}$. The thin, colored lines show modal effective permittivity for an integer mode with harmonic order $\rho \in [1, 60]$. The weighted sum of these modes produces the thick-black line, which is compared to the total effective permittivity obtained from modal expansion simulations (black dots).}
    \label{fig:modal-ereff}
\end{figure}

Considering only the TM components, we recognize that each TM mode has an equivalent capacitance, denoted $C_{in,h}$, and we define a effective capacitance $C_1$ for the sum of TM modes such that
\begin{equation}
    Z_{TM} = \frac{1}{j \omega C_1} = \sum_{h'}^{TM} \frac{A_{h}}{j\omega (C_{in,h}^{(-M)} + C_{in,h}^{(N)})}.
\end{equation}
Dividing both sides by the capacitive impedance of the freestanding metasurface $1/(j\omega C_0)$ yields
\begin{equation}
    \frac{C_0}{C_1} = \sum_{h'}^{TM} \frac{A_{h} C_0}{C_{in,h}^{(-M)} + C_{in,h}^{(N)}}.
    \label{eq:ereff-sum-intermediate}
\end{equation}
The capacitance of an individual mode can be determined from the equation for input impedance of a transmission line, which can be cascaded for each layer in the multi-layer stack-up. Beginning with the transmission line furthest to the right, we have
\begin{equation}
    Z_{\text{in}}^{(2)} = \etah{2} \frac
        {\etah{1} + j\etah{2} \tankzh{2}}
        {\etah{2} + j\etah{1} \tankzh{2}}
        \label{eq:transmission-line-single}
\end{equation}
If we define $C_{0,h}$ as the modal capacitance for the freestanding metasurface, then
\begin{equation}
    \etah{i} = \frac{1}{j\omega \epsilon_r^{(i)}C_{0,h}}
\end{equation}
and \eqref{eq:transmission-line-single} can be written in terms of layer permittivity:
\begin{equation}
    \frac{C_{in,h}^{(2)}}{C_{0,h}} = \epsr{2} \frac
        {\epsr{1} + j \epsr{2} \tanad{2}}
        {\epsr{2} + j \epsr{1} \tanad{2}}.
\end{equation}
 The relationship can be expanded with the identity
\begin{equation}
    j\tan x = \frac{1 - e^{-j2x}}{1 + e^{-j2x}}
\end{equation}
and defining the quantity
\begin{equation}
    r = \frac{\epsr{2} - \epsr{1}}{\epsr{2} + \epsr{1}}
\end{equation}
to obtain the relationship
\begin{equation}
    \frac{C_{in,h}^{(2)}}{C_{0,h}} = \epsr{1} + \left(\epsr{2} - \epsr{1}\right)\frac
        {1-e^{-2\alpha_{h} d_2}}
        {1+re^{-2\alpha_{h} d_2}}
\end{equation}
This is the \textit{effective permittivity} seen by an individual mode, looking into the second layer, and represents a logistic function with respect to the thickness parameter $d_2$; for decreasing thickness the effective permittivity converges to $\epsr{1}$, and for increasing thickness to $\epsr{2}$. It may be further generalized to multiple layers with the recursive relationship:
\begin{equation}
    \erinh^{(i)} = \begin{cases}
        \erinh^{(i-1)} + \left(\epsr{i} - \erinh^{(i-1)}\right)\dfrac
            {1-e^{-2\alpha_{h} d_i}}
            {1+r_i e^{-2\alpha_{h} d_i}}, & i > 1 \\
        \epsr{1}, & i = 1.
    \end{cases}\label{eq:ereff-mode}
\end{equation}
Incorporating $C_{in,h}^{(N)} = \ereff^{(N)} C_{0,h}$ into \eqref{eq:ereff-sum-intermediate} and recognizing that $C_1/C_0$ is the total effective permittivity yields
\begin{equation}
    \frac{1}{\ereff} = \sum_{h'}^{TM} \frac{A_{h} C_0}{2C_{0,h}} \left(\frac{2}{\erinh^{(-M)} + \erinh^{(N)}}\right).
\end{equation}
The convergence rate of each in this sum is shown and compared to the overall convergence in \autoref{fig:modal-ereff}.

A final simplification notes that as the thickness of all dielectric layers approaches zero, the modal effective permittivity approaches the permittivity of the outermost layer. Assuming that the outermost layers in the stack-up are free space (both $\epsr{1}$ and $\epsr{-1}$ are unity), then $\ereff$ must equal one as well. Defining a new modal coefficient
\begin{equation}
    a_h = \frac{A_h C_0}{2C_{0,h}}
\end{equation}
then
\begin{equation}
    \frac{1}{\ereff} = \sum_{h'}^{TM} \frac{2 a_h}{\left(\erinh^{(-M)} + \erinh^{(N)}\right)}
    \label{eq:ereff-sum}
\end{equation}
where
\begin{equation}
    \sum_{h'}^{TM} a_h = 1
\end{equation}

The combination of \eqref{eq:ereff-mode} and \eqref{eq:ereff-sum} forms a rigorous representation of the effective permittivity of a metasurface layer embedded in a set of dielectric layers.

\subsection{Approximation of the infinite summation}

Without prior knowledge of the magnitudes of every harmonic, or at the very least a large number of them, the usefulness of the effective permittivity relation of \eqref{eq:ereff-sum} in prediction of metasurface behavior is quite limited. Ideally, an expression of effective permittivity could be fully described by only a few terms, which would require truncation of the infinite sum in some way. However, the infinite series of harmonics is generally a slowly decaying one, especially for resonant structures, such that even in the best case scenario truncation of the series would yield an expression with several dozen coefficients. 

\begin{figure}
    \includegraphics[center,width=3.49in]{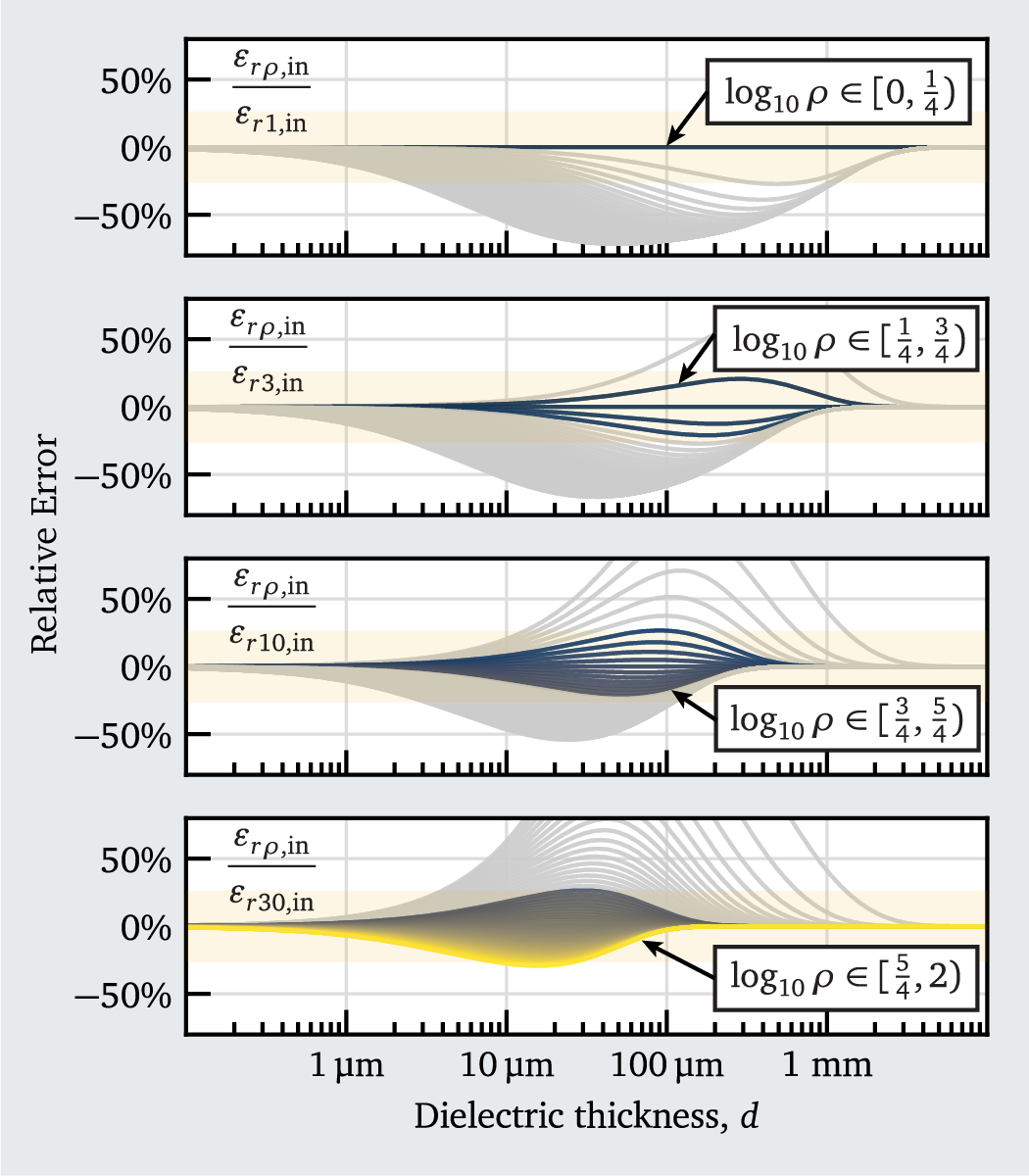}
    \caption{Relative error between approximating modes $\rho_k \in \{1, 3, 10, 30\}$ and partitions of neighboring modes. Colored lines indicate the relative error of modes in the neighborhood of $\rho_k$, while the light gray lines show the error of integer modes in $\rho \in [1, 60]$. }
    \label{fig:ereff-harmonic-error}
\end{figure}

Instead, we take advantage of the similarity of neighboring harmonics, and the fact the rate of evanescent decay smoothly and gradually increases with increasing harmonic order. First, we note all harmonics with the same harmonic radius $\rho_{mn}$ share identical effective permittivity behavior, where
\begin{equation}
    \rho_{mn} = \sqrt{m^2 \left(\frac{P_y}{P_x}\right) + n^2 \left(\frac{P_x}{P_y}\right)}.
\end{equation}
We suggest that the infinite set of harmonics can be partitioned into sets of harmonics based on the magnitude of the harmonic order $\rho$, so that the sum of all harmonics in a particular range $[\rho_0, \rho_1]$ can be approximated by a single harmonic; that is, 
\newcommand{\erinph}{\epsilon_{r\rho_k,\text{in}}}
\begin{equation}
\frac{1}{\ereff} \approx \sum_{k=1}^{K} \frac{2b_k}{\erinph^{(-M)} + \erinph^{(N)}}
\end{equation}
where $B = \{b_1, b_2,\ldots,b_K\}$ are harmonic coefficients with $\sum B = 1$ and $P = \{\rho_1, \rho_2, \ldots, \rho_K\}$ are equally spaced harmonic orders, which we will call \textit{approximating orders}. For this work, we choose $K=4$ and
\begin{equation}
    P = \left\{ 10^{(k-1)/2} : k = \{1,2,\ldots,K\}\right\}
\end{equation}
which is approximately $P \approx \{1, 3, 10, 30\}$. These modes are chosen \textit{a priori} to improve the simplicity of the model, although one could also allow the harmonic order to vary as part of a fitting procedure to better fit the sum of neighboring modes.

The suitability of a single mode as a representation of the weighted sum of many nodes of similar order is shown in \autoref{fig:ereff-harmonic-error}. Here the relative error between the modal effective permittivity described by \eqref{eq:ereff-mode} of a chosen approximating mode $\rho_k \in P$ and the modal effective permittivity of a partition of neighboring modes -- the term \textit{neighboring} describes a logarithmic distance from the chosen mode -- is illustrated. When partitioned in this way, every harmonic can be represented by the approximating harmonic with less than 30\% relative error. While this level of error seems substantial, the harmonic orders in the weighted sum are relatively evenly distributed above and below the approximating order, lowering the error in the weighted mean.


\subsection{Process for obtaining coefficients}

A total of $K$ simulations are required to fully specify the effective permittivity model. Because the effective permittivity is defined as the ratio of a particular case's capacitance to the capacitance of the freestanding case, it is first necessary to model the freestanding case. Then we must perform at least $K-1$ simulations to determine the $K$ coefficients of the model; the $K$-th term is determined by the fact that the sum of coefficients must equal one. Because the behavior of individual modes is primarily driven by thickness, we model each case with the metasurface placed between identical substrates of varying thickness and a single permittivity (in this example, $\varepsilon_r = 3$). 

\begin{figure}
    \centering
    \includegraphics[width=3.49in]{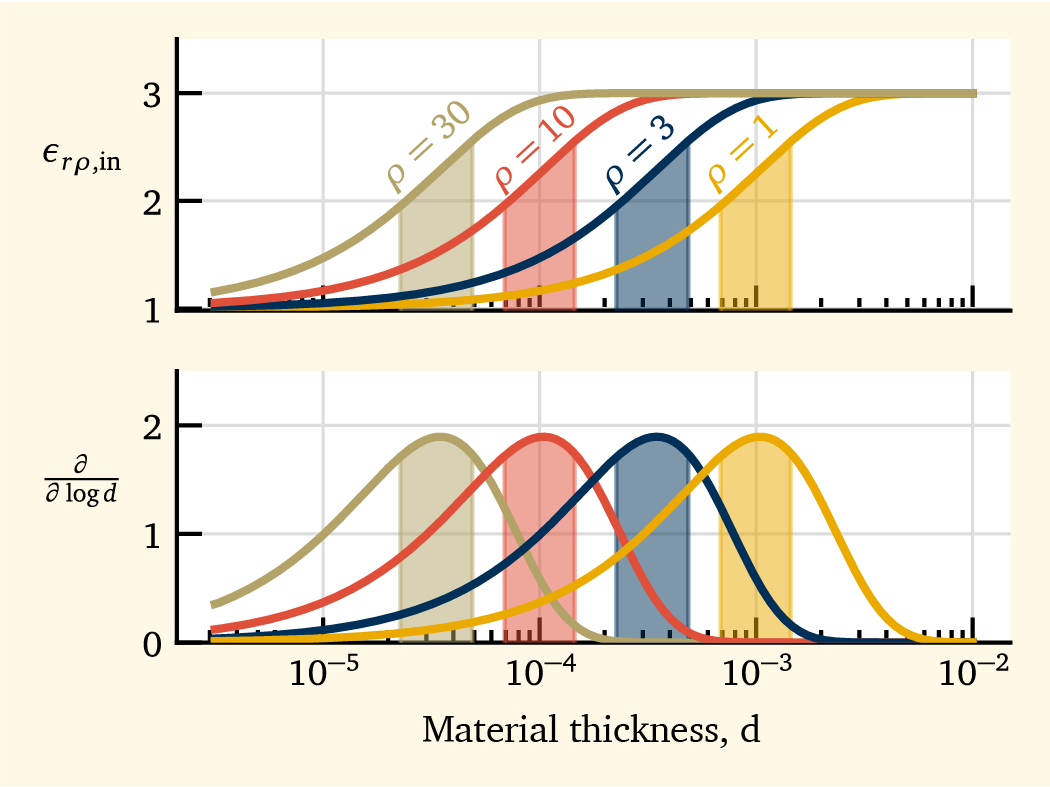}
    \caption{Model coefficients are obtained via nonlinear least squares by simulating the metasurface with different symmetric substrates. The thickness of the substrate in these models is chosen based on sensitivity in the approximating modes. On top, the effective permittivity of mode $\rho$ looking into a substrate of thickness $d$ is shown, with the rate of change with respect to log of thickness below. Sampled thicknesses should be in the ranges shown.}
    \label{fig:thickness-choices}
\end{figure}

Any solver may be employed for this purpose; however, the method of obtaining the capacitance of each model may be varied based on the type of solver chosen. For time-domain methods where many frequency points are available, the capacitance of a metasurface resonance may be obtained through the reactance slope with the relationships of \eqref{eq:L-from-slope} and \eqref{eq:C-from-slope}. For many frequency-domain methods, it may be advantageous to model only a single frequency-point. In this case, the metasurface should be modeled at a low frequency -- at least two decades below the cutoff frequency of the first evanescent mode -- from which reactance $X_{\text{eq}}$ can be computed by \eqref{eq:impedance-from-S11} and capacitance obtained from the reactance by
\begin{equation}
    C = \frac{1}{j\omega X_{\text{eq}}(\omega \to 0)}
\end{equation}
since the contribution to the reactance from any inductance is small at low frequencies.  Dividing the capacitance obtained for each model by the capacitance of the freestanding metasurface yields the effective permittivity for each case. 

The choice of thicknesses to model for obtaining the coefficients is related to the approximating modes of the model. It is important to model thickness where the modal effective permittivity varies with respect to thickness. \autoref{fig:thickness-choices} shows the convergence of modal effective permittivity for the approximating modes previously defined as $P$ and also shows the rate of change of that effective permittivity with respect to the $\log$ of thickness. A good choice of substrate thicknesses is near where the rate of change for each thickness is largest; for a periodicity of $\SI{10}{\mm} $ and $P=\{1, 3, 10, 30\}$, this corresponds to thicknesses of $\SI{30}{\um}$, $\SI{100}{\um}$, $\SI{300}{\um}$, and $\SI{1}{\mm}$.

Nonlinear least squares \cite{more1978} is used to compute the coefficients of the effective permittivity model from the $K - 1$ effective permittivity data points produced by simulation. 

\subsection{Modeling results}

\begin{figure}
    \centering
    \includegraphics[width=3.49in]{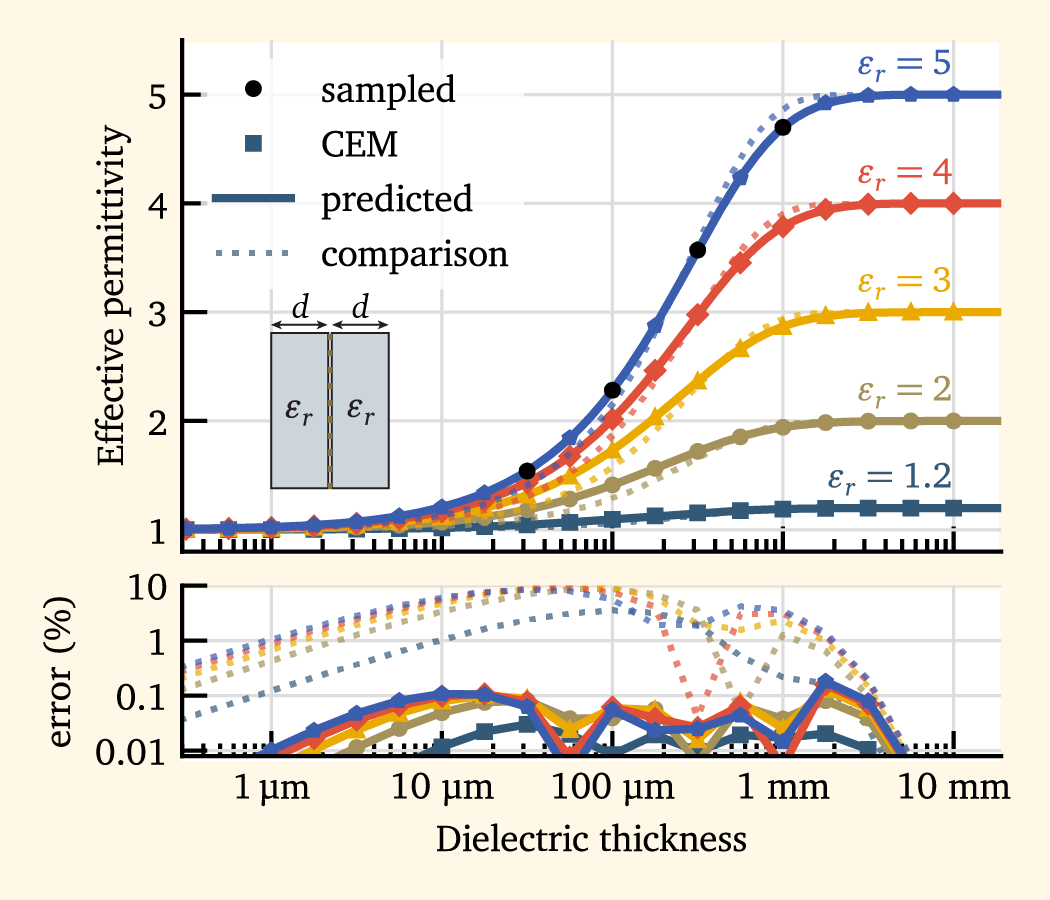}
    \caption{Effective permittivity versus thickness for a symmetric pair of dielectric layers of permittivity $\epsilon_r$ and thickness $d$. Each marker represents a separate simulation, which is compared to predictions from the single-term effective permittivity model of \cite{costa2021} (dashed lines) and the model from this work (solid lines). The relative error between each and the CEM cases are shown below. The black dots represent simulations sampled to train the model.}
    \label{fig:ereff-twoside}
\end{figure}

The predictive capability of the effective permittivity model is illustrated in \autoref{fig:ereff-twoside}. From four simulations of the dipole metasurface, coefficients of $b_k = \{0.109, 0.421, 0.358, 0.112\}$ are obtained from the least squares fit. This model is then used to predict the effective permittivity for dielectric layers ranging from $\varepsilon_r = 1.2$ to $\varepsilon=5$ with thicknesses between $\SI{0.1}{\um}$ and $\SI{10}{\mm}$. When compared to results from CEM, the predicted effective permittivity has less than 0.2\% error over the entire range of thicknesses and permittivities, compared to 9.6\% error with the single-term model of \cite{costa2021}, nearly two orders of magnitude improvement in relative error.

\begin{figure}
    \centering
    \includegraphics[width=3.49in]{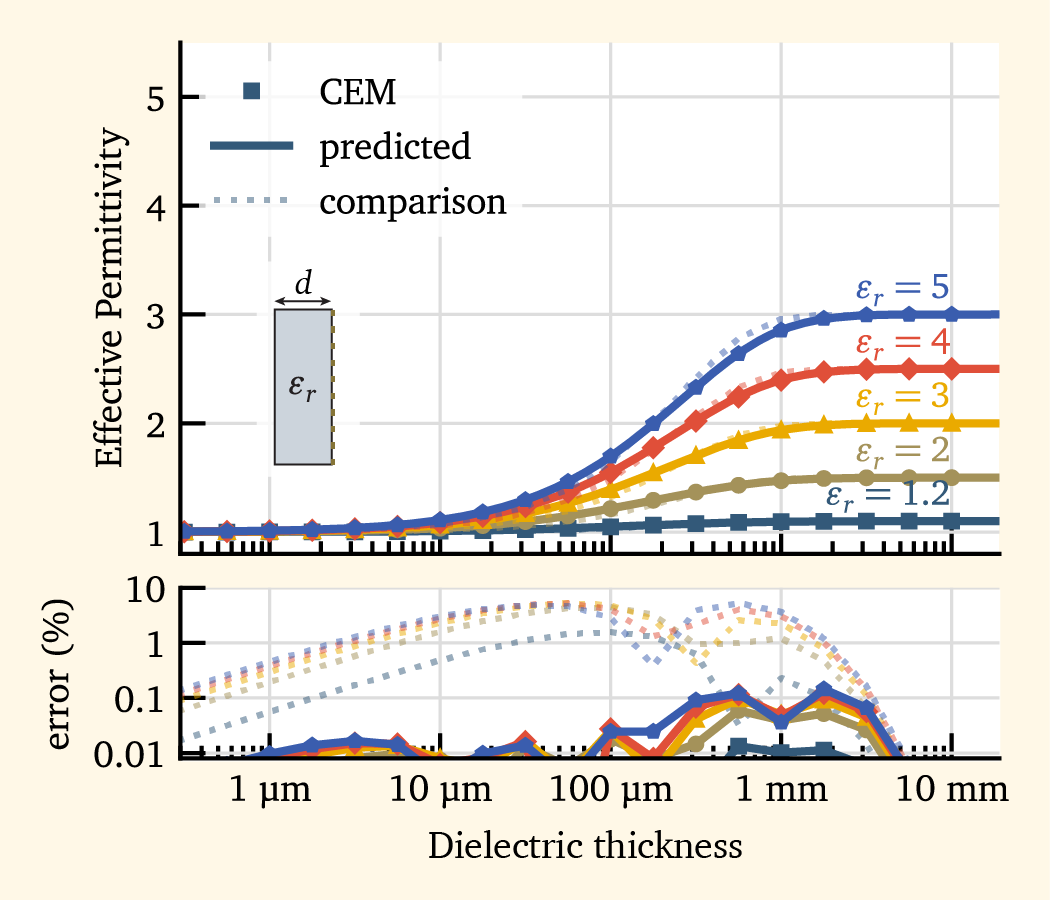}
    \caption{Prediction of effective permittivity versus thickness for a single dielectric layer to the left of the metasurface with permittivity $\epsilon_r$ and thickness $d$. The metasurface is identical to that of \autoref{fig:ereff-twoside}.}
    \label{fig:ereff-oneside}
\end{figure}

Similarly, \autoref{fig:ereff-oneside} shows predictions of the same model for a single substrate on one side of the metasurface with varying thickness. Once again, the relative error over the entire set of simulations is less than 0.2\%.

\section{Combination of Modal Equivalent Circuit and Modal Effective Permittivity Model}

It may be supposed that for a metasurface with an LC resonant response, the determination of effective permittivity for a layered dielectric is all that is needed to fully characterize change in the metasurface response from the freestanding case. However, this is predicated on two related assumptions: first, that the values of the effective capacitance and inductance remain constant with frequency; and second, that the layered dielectric environment has no effect on the effective inductance of the structures. As we will now explore, the presence of dispersion in the evanescent modes, as shown in \autoref{fig:harmonic-lumped}, stresses the validity of these assumptions.

\subsection{Dispersion due to higher order modes}

\begin{figure*}
    \centering
    \includegraphics[width=7.14in]{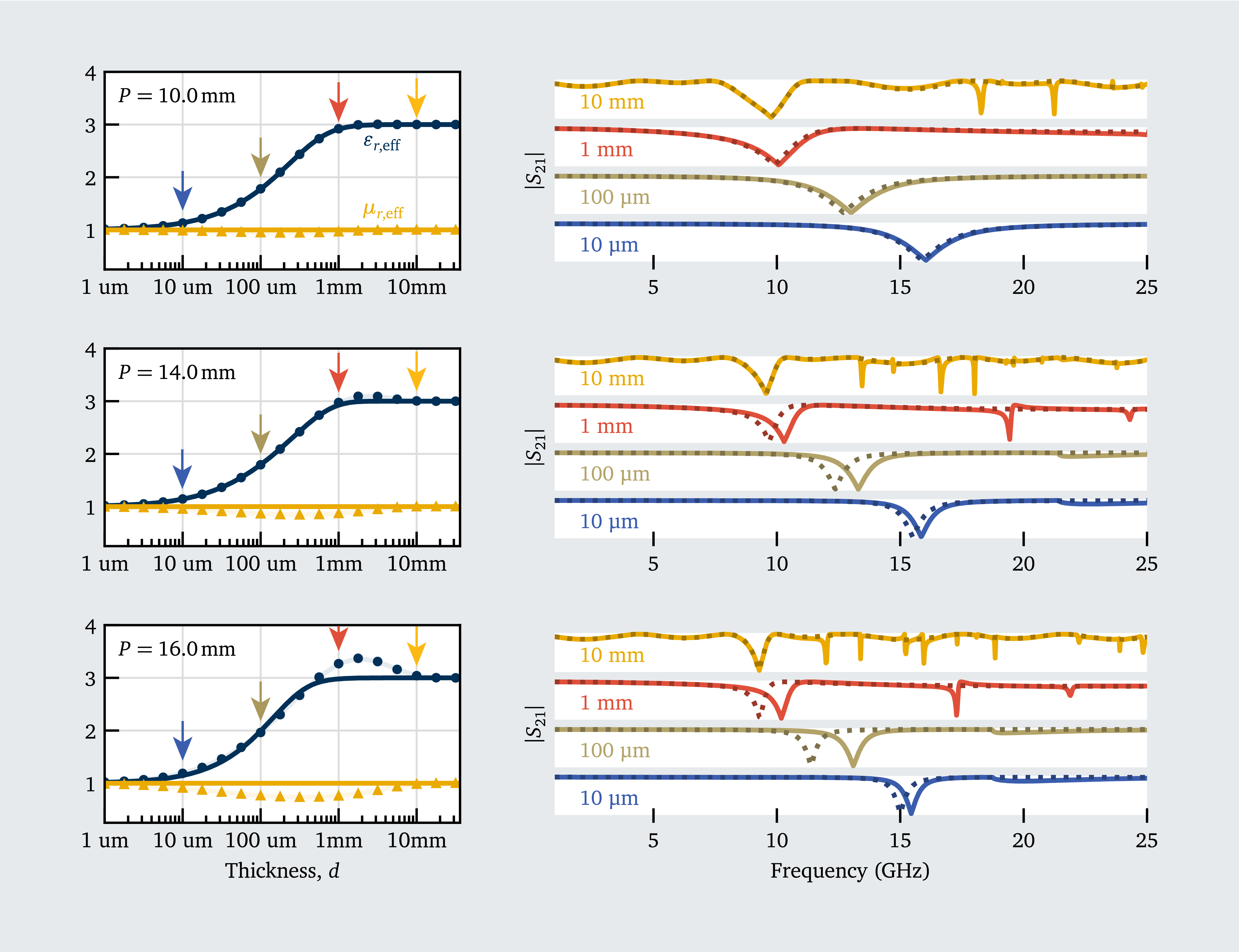}
    \caption{On left, effective permittivity extracted from modal expansion simulations (dots) compared to prediction by the effective permittivity model (lines) at three different periodicities $P$ (top to bottom). On right, qualitative comparisons of frequency response between the modal expansion calculation (lines) and the effective permittivity model (dashed lines) at four different dielectric thicknesses, assuming $\ureff=1$.}
    \label{fig:hmodel-bad}
\end{figure*}

We have up to this point relied on treating each evanescent mode as having an equivalent static capacitance or inductance. For the vast majority of higher order modes, this assumption is reasonable. However, for the lowest order modes, especially for periodic surfaces with a large periodicity, the capacitance varies considerably with respect to frequency, causing an additional source of variation in the effective permittivity. This dispersion also appears in the TE modes, causing the effective inductance of the structure to vary. In this way it is possible for layers composed only of dielectrics to present an \textit{effective permeability} to the metasurface. These effects, and the inability of the effective permittivity model to capture them, are shown in \autoref{fig:hmodel-bad}. 

At the smallest periodicity $P=\SI{10}{\mm}$, the effective permittivity model performs well; there is less than $0.2\%$ error in agreement between $\ereff$ extracted from rigorous solutions and that produced by the model. Even so, small deviation in resonant frequencies occur at intermediate thicknesses of substrate in spite of the good fit of the effective permittivity relationship. This is caused by a non-unity effective permeability, which is the equivalent inductance of the metasurface as material thickness changes divided by the freestanding surface inductance. At its greatest point of variation, the effective permeability of the $P = \SI{10}{\mm}$ dipole array is 0.94.

The effect is even more pronounced at higher periodicity, where $\ureff=0.844$ at $P=\SI{14}{\mm}$ and $\ureff = 0.736$ at $P=\SI{16}{\mm}$ can be found. Furthermore, the effective permittivity relationship is no longer monotonically increasing, overshooting the substrate permittivity value before settling back down to it. An additional consequence of the simple LC equivalent circuit model of \eqref{eq:simple-LC} is that higher frequency behavior approaching cutoff frequencies of the $TE_{10}$ and $TM_{01}$ modes and beyond is not accurately modeled. 

These problems can be addressed by the use of an equivalent circuit model based on the rigorous modal expansion of \eqref{eq:Zeq-sum} and outlined in \cite{garcia-vigueras2012} and \cite{rodriguez-berral2015}, namely that
\begin{align}
    \begin{split}
    \Zeq ={}& \sum_{h}^{N_{TE}} A_{hTE} Z_{in,h}
    + \sum_{h}^{N_{TM}} A_{hTM} Z_{in,h}\\
    &+ \frac{1}{j\omega C_{ho}} + j\omega L_{ho}
    \end{split}
\end{align}
where $Z_{in, h}$ is the parallel combination of modal impedances looking into the dielectrics layers to the left and right of the metasurface, as defined by \eqref{eq:transmission-line}. With this formulation, the impact of dielectric layers on $N_{TE} + N_{TM}$ modes considered to be dispersive can be calculated analytically, while the higher-order capacitance $C_{ho}$ can be modified by the effective permittivity model.


Setting $N_{TE} = N_{TM} = 1$ is sufficient to capture the dispersive effects, although in theory additional modes could be included to extend the frequency range of validity of the model. The lowest order modes are the $TE_{10}$ and $TM_{01}$ modes, and therefore the frequency response of the freestanding metasurface can be fully specified by four unknowns: modal coefficients $A_{TE_{10}}$ and $A_{TM_{01}}$, and higher-order circuit values $L_{ho}$ and $C_{ho}$. It is our objective to discover the appropriate values of each of these, as well as an effective permittivity model for $C_{ho}$ which is capable of predicting the frequency response for any set of dielectric layers.

\newcommand{\ATE}{A_{\mathrm{TE_{10}}}}
\newcommand{\ATM}{A_{\mathrm{TM_{01}}}}

We begin with initial guesses for $\ATM$ and $\ATE$, which will eventually be varied by a least squares optimization procedure. Values for the higher-order $L_{ho}$ and $C_{ho}$ may then be obtained from a simulation of the freestanding metasurface. The capacitive components $C_{ho}$ and $\eta_{\mathrm{TM_{01}}}$ and inductive components $L_{ho}$ and $\eta_{\mathrm{TE_{10}}}$ may be considered in turn by sampling the freestanding metasurface simulation at some low frequency ($\omega_{\ell} = \sim 0.01\omega_c$) and a high frequency ($\omega_{h} = \sim 0.98\omega_c$) where $\omega_c$ is the cutoff frequency of the lowest-order mode. The total capacitance of the surface, which due to the higher order modes now varies with frequency,  is calculated from simulated impedance
\begin{equation}
    C_0 = \operatorname{Re}\left\{\frac{1}{j\omega \Zeq(\omega_\ell)}\right\}
\end{equation}
and modal capacitance for the $\mathrm{TE_{10}}$ mode calculated similarly
\begin{equation}
    C_{TM_{01}} = \operatorname{Re}\left\{\frac{1}{j\omega \eta_{\mathrm{TM_{01}}}(\omega_\ell)}\right\}
\end{equation}
from which it is possible to calculate the higher-order capacitance
\begin{equation}
    C_{ho} = \left[\frac{1}{C_0} - \frac{\ATM}{C_{TM_{01}}}\right]^{-1}.
\end{equation}

\begin{figure*}
    \centering
    \includegraphics[width=7.14in]{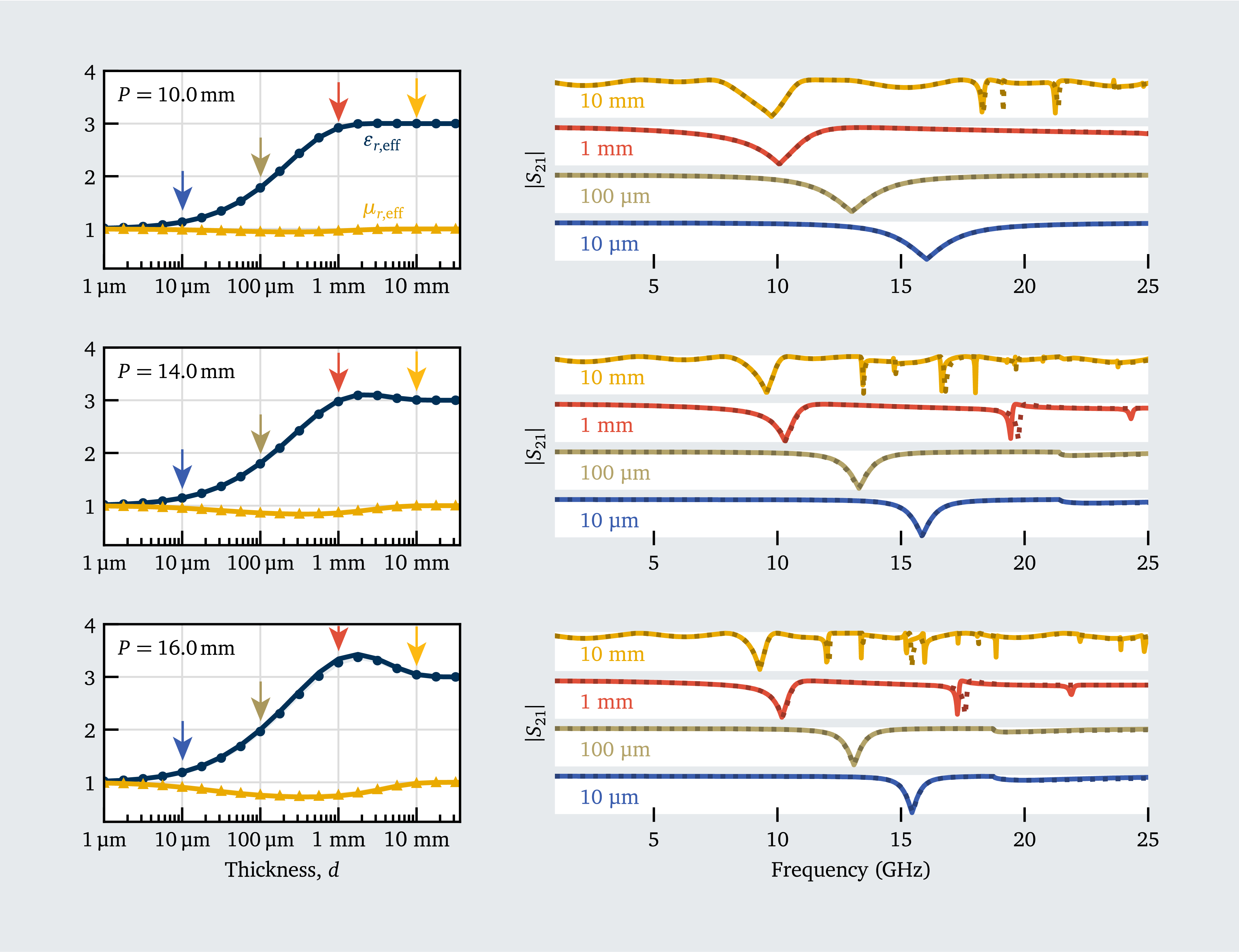}
    \caption{Unlike the effective permittivity model alone as shown in \autoref{fig:hmodel-bad}, the higher-order harmonic model accurately predicts the impact from modal dispersion on equivalent inductance capacitance (left), reproducing nearly the same frequency response of the modal expansion (comparison on right; solid lines are modal expansion model, dashed lines are the higher-order model).}
    \label{fig:hmodel-fixed}
\end{figure*}

We now turn to consideration of data sampled at the higher frequency to determine $L_{ho}$. We first subtract from $\Zeq$ the effect of the higher-order capacitance and dispersive TM mode:
\begin{equation}
    Z_L = \left.\Zeq - \left(\frac{1}{j\omega C_{ho}} + A_{TM_{01}} \eta_{TM_{01}(\omega)}\right)\right|_{\omega=\omega_{h}}.
\end{equation}
The equivalent inductive of  $Z_L$ and the modal TE impedance $Z_L$ are calculated as
\begin{align}
    L_0 &= \operatorname{Re}\left\{\frac{Z_L(\omega_h)}{j\omega_h}\right\} \\
    L_{\mathrm{TE_{10}}} &= \operatorname{Re}\left\{\frac{\eta_{\mathrm{TE_{10}}}(\omega_h)}{j\omega_h}\right\}
\end{align}
and the higher-order inductance is computed to be
\begin{equation}
    L_{ho} = L_0 - \ATE L_{\mathrm{TE_{10}}}.
\end{equation}
We are left with estimates for $L_{ho}$ and $C_{ho}$ based on some guess of $\ATM$ and $\ATE$; what remains is to determine the correct values for these terms, as well as the coefficients of the modal effective permittivity model which is to properly scale the $C_{ho}$ term when predicting the response from a layered dielectric environment. Nonlinear least squares may be used for this purpose, in a similar manner as for the effective permittivity model. In fact, the same simulations of the metasurface with various thicknesses of dielectric used to produce coefficients of the modal effective permittivity model may be used to fit the $\ATM$ and $\ATE$ terms, with an additional frequency point sampled somewhere between $\omega_\ell$ and $\omega_h$ at each thickness. The best estimate for $\ATM$ and $\ATE$ will most closely reproduce these mid-band data points -- the points at $\omega_\ell$ and $\omega_h$ are pinned by the solution for $L_{ho}$ and $C_{ho}$ -- and will also produce a value for $L_{ho}$ that is constant with respect to both dielectric thickness and frequency. 

\begin{table}
    \centering
    \caption{Modal coefficients of higher-order model}
    \label{tbl:hmodel-coeffs}
    \renewcommand{\arraystretch}{1.3}
    \begin{tabularx}{1.0\linewidth}{Xccccc}\toprule
    Period & $\ATM$ & $\ATE$ 
        & $L_{ho}$ & $C_{ho}$
        & $\ereff$ model\\
    & & & (nH) & (pF) & coefficients \\\midrule
    \SI{10}{\mm} & 0.68 & 1.08
        & 5.59 & 12.0
        & 0.0, 0.46, 0.42, 0.12 \\
    \SI{14}{\mm} & 0.60 & 1.05 
        & 12.9 & 5.85
        & 0.0, 0.37, 0.42, 0.22\\
    \SI{16}{\mm} & 0.15 & 1.06 
        & 17.9 & 4.25
        & 0.01, 0.36, 0.36, 0.27\\
    \bottomrule
    \end{tabularx}
\end{table}

The complete set of model terms calculated for the dipole metasurface arrays have been summarized in \autoref{tbl:hmodel-coeffs}, for each of the three periods previously highlighted in \autoref{fig:hmodel-bad}. Restating the significance of these coefficients, these eight terms are capable of accurately predicting the frequency response of the dipole metasurface over the entire frequency range up to cutoff of the first evanescent mode, for any arbitrary layered dielectric environment on either side of it, in only a few milliseconds.

\autoref{fig:hmodel-fixed} shows the same cases as \autoref{fig:hmodel-bad}, but using the higher-order model coupled to the effective permittivity model instead of the effective permittivity model alone. By including the variation in frequency of the two lowest-order modes, the higher-order model is able to  predict both the non-monotonic behavior of the effective permittivity and the variation in effective permeability which occurs at higher cell periods. An additional benefit is that the frequency response is accurately modeled past the first diffractive order, unlike the simple equivalent circuit model which assumes that the frequencies of interest are far below the cutoff frequency for any higher-order harmonics.

\section{Example Applications}

The method described is general enough to be applied to a large variety of problems; however, we note two immediate applications.

\subsection{Accurate modeling of bonding layers and solder mask in multilayer PCBs}

It is often the case that embedding a metasurface or frequency-selective surface in symmetric dielectric layers improves the angular stability of the response \cite{munk2000}. When fabricated for microwave frequencies by either printed circuit board technologies or composite layup techniques, practical fabrication considerations require a bonding layer or pre-preg film to be placed between the metal and the second substrate. This bond film may have different electromagnetic properties from the substrate, and is usually deeply sub-wavelength and a fraction of the thickness of the substrate. It may be tempting to eliminate this bonding layer in the CEM model because of its small thickness, but we must remember that strong evanescent fields immediately adjacent to a periodic surface have a high level of interaction with the region of space nearest to them. It is important to account for even the thinnest of layers when they are adjacent to an impedance discontinuity.

Not every full-wave modeling scheme is amenable to inclusion of such layers; for example, an FDTD simulation may have an otherwise reasonable discretization that is larger than the thickness of the bond layer, and increasing the FDTD grid granularity to accommodate the layer carries with it a large penalty in computational time. A \SI{0.1}{\mm} thick layer in a metasurface model at a maximum frequency of 10 GHz would require a discretization of $\Delta = \lambda/300$; when compared to a typical discretization of $\Delta = \lambda/20$, use of a fine grid to capture the effects of the thin layer would take $15^4 = \SI{50625}{}$ times longer to solve. Even a conservative model at $\Delta = \lambda / 100$ would require an 81 times penalty to incorporate the thin layer. By contrast, the effective permittivity model can be solved using five of the coarse simulations. A meshed finite element solver may also suffer from a large increase in mesh density away from the thin layer in order to accommodate it. Here the effective permittivity model offers an alternative, flexible approach.

\begin{figure}
    \centering
    \includegraphics[width=3.49in]{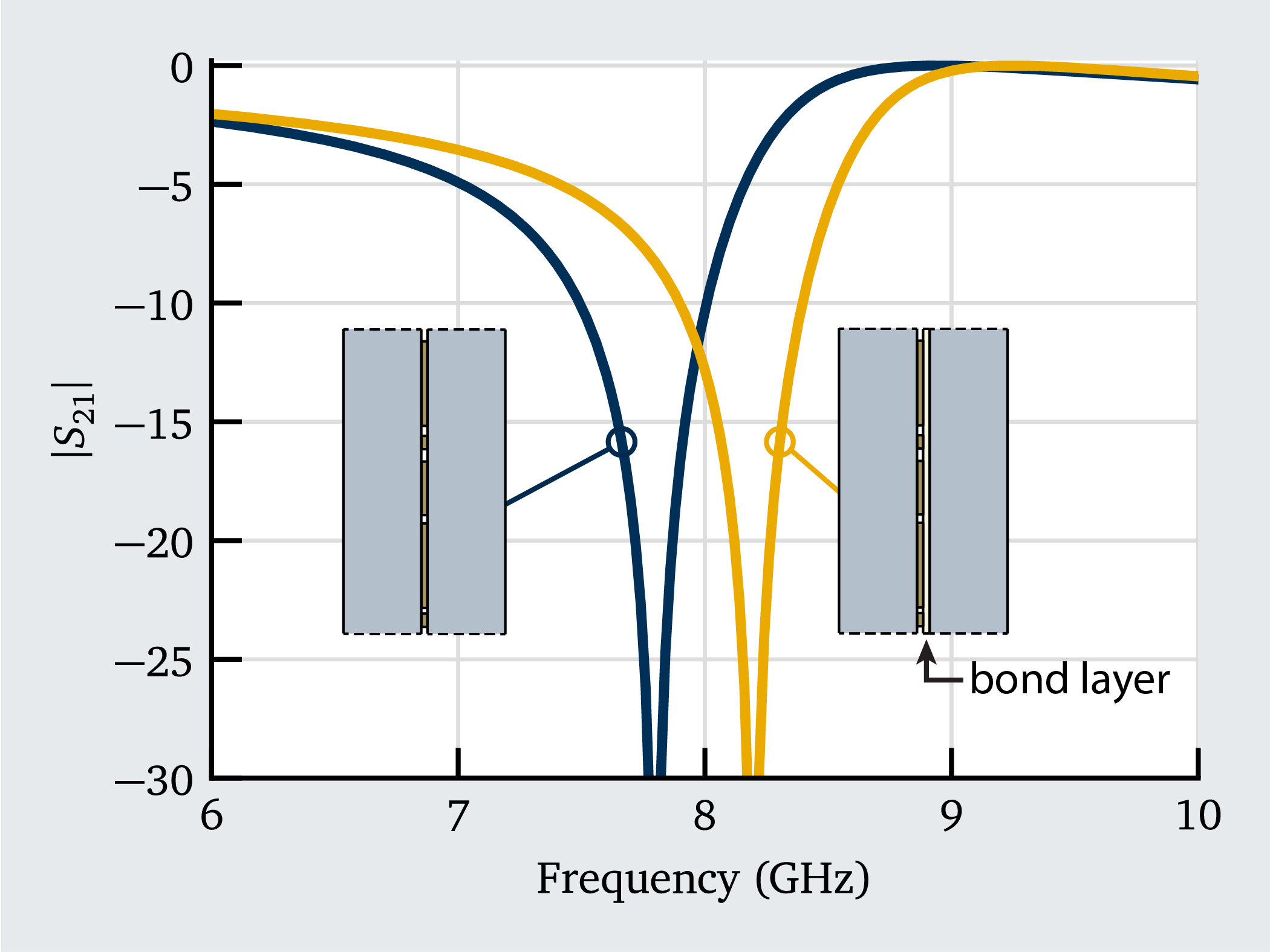}
    \caption{Transmission vs. frequency for a dipole metasurface embedded between two layers of $\varepsilon_r = 6.0$ material with thickness $d=\SI{3.0}{\mm}$. The curve on the right illustrates a 6\% resonance shift that occurs when a \SI{0.076}{\mm} thick bond layer ($\epsilon_r=2.9$) is introduced.}
    \label{fig:bond-layer-impact}
\end{figure}

After modeling a sufficient number of cases to obtain the coefficients for the modal effective permittivity model, the analytic model can then be used to rapidly explore the impact of various bond film permittivities and thicknesses. An example is shown in \autoref{fig:bond-layer-impact}; a metasurface is embedded in symmetric layers of $\varepsilon_r=6$ material with a thickness of $\SI{3}{\mm}$. With the modal effective permittivity, inserting a bond layer of $\epsilon_r=2.9$ and only $\SI{0.076}{\mm}$ is demonstrated to perturb the resonance of the metasurface by 6\%. 

\begin{figure}
    \centering
    \includegraphics[width=3.49in]{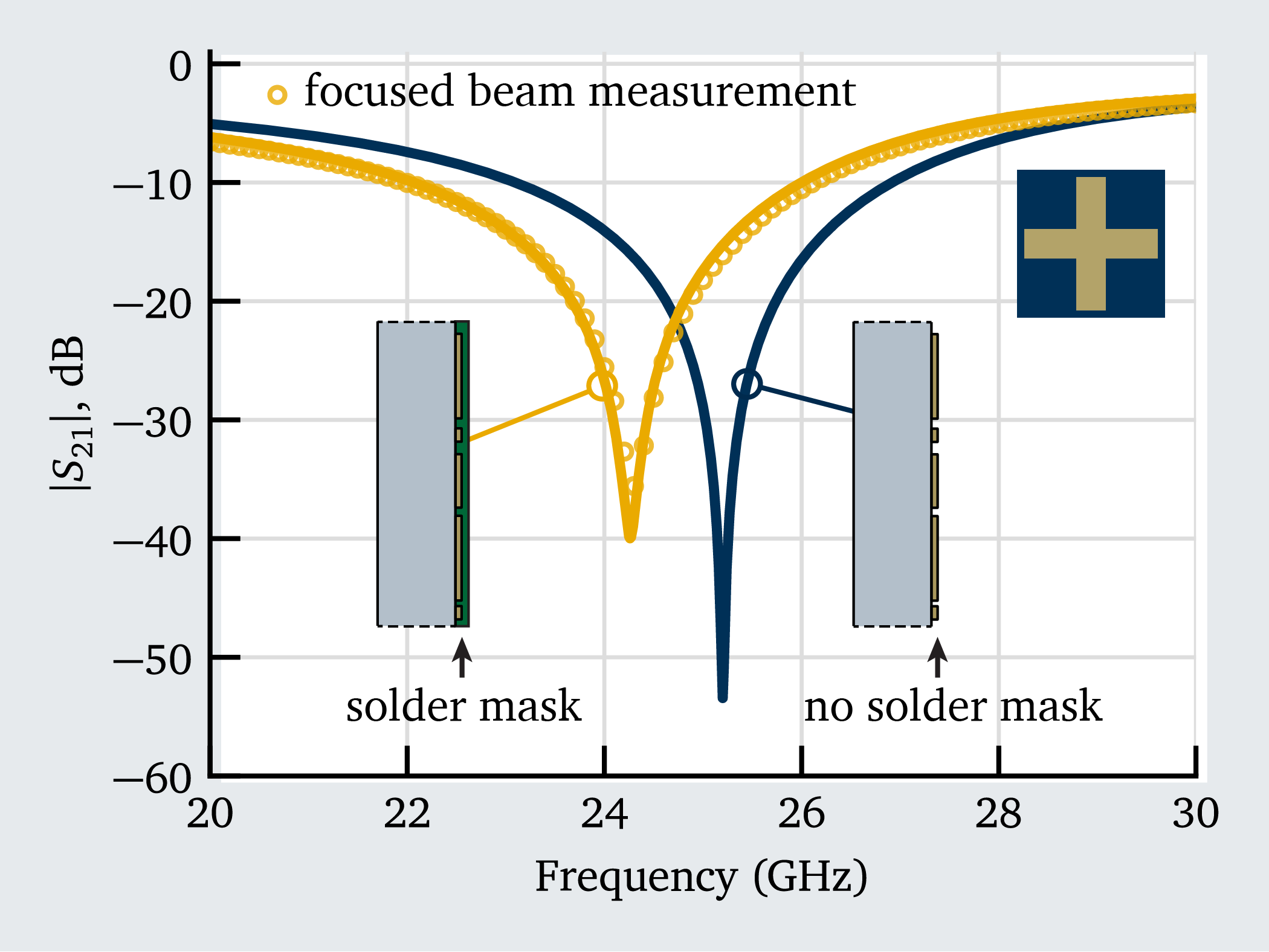}
    \caption{Transmission through a crossed dipole surface (\SI{5.08}{\mm} period, \SI{4.57}{\mm} length, \SI{1.02}{\mm} width on a
    \SI{1.52}{\mm} AD255C-IM substrate. The effective permittivity model may be used to calculate the effects of a very thin ($\SI{25}{\um}$) solder mask layer ($\varepsilon_r = 3.5$, $\tan\delta = 0.045$), which closely matches the transmission measured with the
    focused beam method.}
    \label{fig:soldermask-impact}
\end{figure}

Solder mask on an exterior metasurface layer can also affect the metasurface response. This impact is evaluated for a cross dipole FSS in \autoref{fig:soldermask-impact}, which compares the transmission response of an FDTD model of the FSS without any solder mask layer to the modal effective permittivity model which includes a \SI{25}{\um} solder mask layer ($\epsilon_r=3.5$, $\tan\delta=0.045$ \cite{barnes2005}). The FSS, which has a period of \SI{5.08}{\mm} and crossed dipoles of \SI{4.57}{\mm} length and \SI{1.02}{\mm} width, was fabricated on \SI{1.52}{\mm} Rogers AD255C-IM substrate ($\epsilon_r=2.6$, $\tan\delta=0.0013$ \cite{rogerscorporation2023}), and measured in the focused beam system according to the method outlined in \cite{howard2022b}. The transmission response predicted by the effective permittivity model matches much more closely with the measured data than the full-wave model where solder mask is ignored, and even predicts the effects of loss in the dispersive solder mask material.

\subsection{Loss tangent measurement surface}

The resonance shift caused by dielectric layers as described throughout this work allows metasurfaces to be used for sensing the complex permittivity of dielectrics, as in the loss tangent measurement surface (LTMS) described in \cite{howard2022b}. This measurement technique uses characterization of the center frequency and Q-factor of a resonance in a metasurface to obtain estimates for real permittivity and loss tangent. A significant challenge in the technical implementation of an LTMS is the lack of an analytic relationship between material properties and the measured resonance. A limited matrix of dielectric constant and loss tangent values can be evaluated in a full wave solver to obtain discrete samples of the relationship, but even this quickly becomes intractable when one considers that practical use of the technique must support the characterization of a material of any arbitrary thickness. 

\begin{figure}
    \centering
    \includegraphics[width=3.49in]{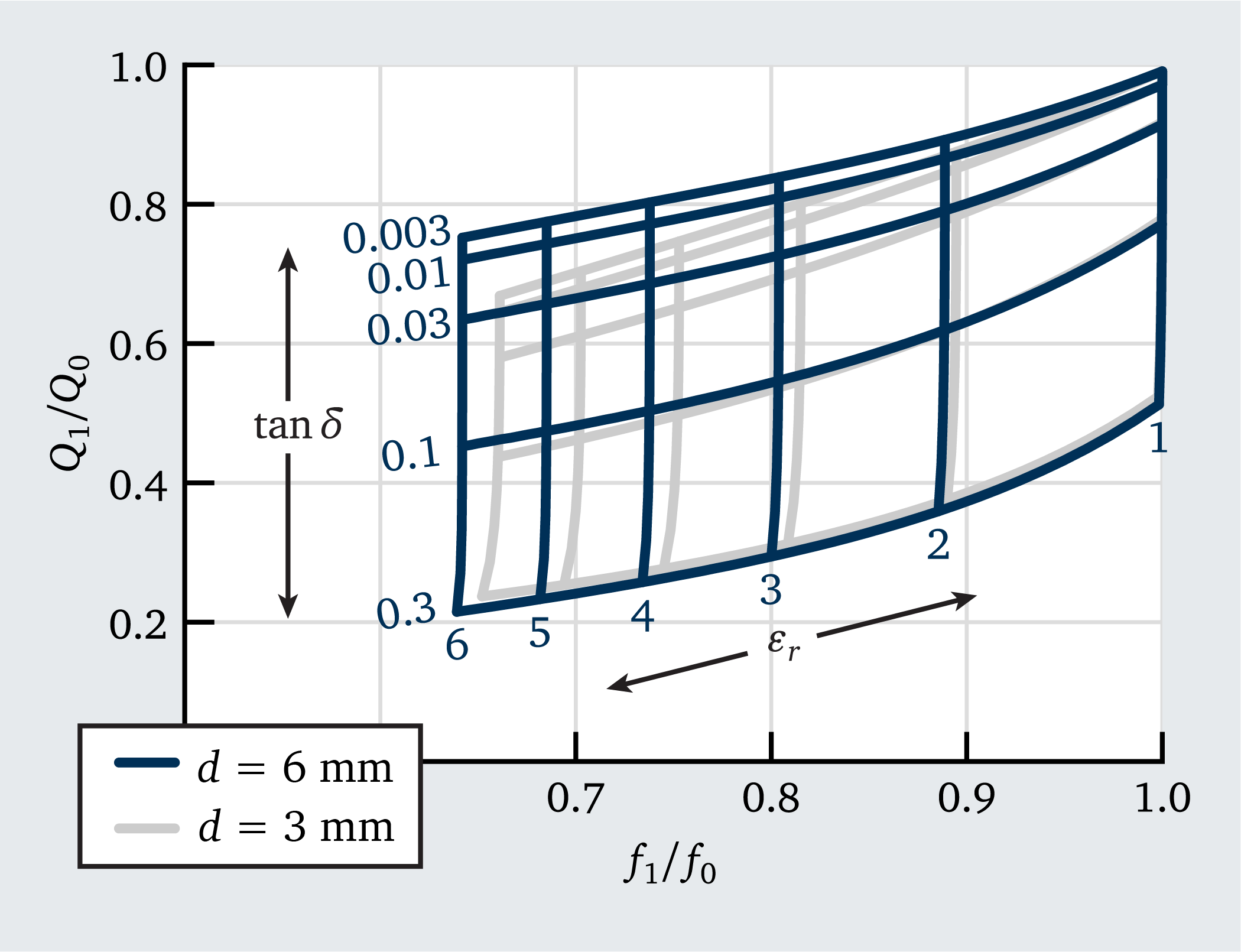}
    \caption{The effective permittivity model can be used to rapidly compute a mapping of resonant frequency shift ($f_1 / f_0$) and change in resonance Q-factor ($Q_1 / Q_0$) onto a determination of real permittivity ($\varepsilon_r$) and loss tangent ($\tan\delta$) of a material under test, as in a loss tangent measurement surface (LTMS). The thickness of the material $d$ has a substantial impact on the resulting mapping.}
    \label{fig:ltms-mapping}
\end{figure}

This problem can be efficiently handled by the effective permittivity model in this work; once the effective permittivity model coefficients are obtained from a few coarse full-wave models as described in previous sections, a mapping of complex permittivity onto frequency and Q-factor can be produced for any arbitrary thickness in a matter of seconds. Such a mapping is shown in \autoref{fig:ltms-mapping} for real permittivities between $\varepsilon_r = 1$ and $\varepsilon_r = 6$, and loss tangents from $0.003$ to $0.3$. Separate curves for thicknesses of $d=\SI{3}{\mm}$ and $d=\SI{6}{\mm}$ are shown to highlight the variation in the effective permittivity relationship with respect to thickness, not only for the real permittivity but also for the loss in the material under test. The accurate determination of the complex permittivity of a material with a metasurface resonator, especially the loss tangent, is seen to require consideration of the material thickness; at a real material permittivity $\varepsilon_r=6$, a change in Q-factor of 0.65 represents nearly an order of magnitude difference in material loss tangent between the two thicknesses.

\section{Conclusion}

Through the development of a robust effective permittivity model in this work based on the modal expansion of fields on a periodic surface, a few key observations have arisen. First, a robust effective permittivity model must include the effects of multiple harmonics of varying evanescent decay, in contrast to previous models which have focused on only a single term. Second, despite the fact that a metasurface is embedded in multiple layers of pure dielectric, it is nevertheless possible for the layers to present an effective permeability to the metasurface, varying its equivalent inductance. This effect, which is more pronounced at larger metasurface periods, has to the best of our knowledge not been previously explored in other effective permittivity works. One must take care to separate changes in capacitance from inductance to fully characterize any resonances produced by the metasurface. We show that dispersion in the lower order modes must be accounted for to properly reproduce these effects in an analytical model. Finally, a model which incorporates these characteristics is a predictive one; once the coefficients for the model have been determined, it is capable of producing analytically and with great accuracy the frequency response for any potential configuration of dielectric layers in only a fraction of a second.

Several opportunities remain for continued development of the model. This work has demonstrated the application of the effective permittivity model to a simple dipole metasurface, but we anticipate that it can readily be applied to any number of other metasurface types of varying complexity, and intend to demonstrate its application to an advanced metasurface in a future work. While both the effective permittivity model and the higher-order equivalent circuit model are easily applicable to the metasurfaces which contain multiple resonances, the procedure described in this work to compute coefficients for the model made particular assumptions about the nature of the single resonance present, so a more general method is required for obtaining the coefficients. 

\section*{Acknowledgment}

The authors thank Dr. R. Todd Lee for introducing the modal expansion concept to Mr. Howard and Dr. David R. Reid for fruitful discussions on the topic. This work was supported by the GTRI's Independent Research and Development portfolio.

\bibliographystyle{IEEEtran}
\bibliography{references}

\end{document}